\documentclass[a4paper]{raa}            % referee version: for submission
\usepackage{graphicx,times,amsmath,hyperref,tabularx}             %for PS/EPS graphics inclusion, new

\begin{document}

   \title{Mapping the Milky Way with LAMOST I: Method and overview}
%   \subtitle{I. Place Your Subtitle Here}

   \volnopage{Vol.0 (200x) No.0, 000--000}      %%preserved for Editor. DOn't remove!
   \setcounter{page}{1}          %%starting page, preserved for Editor. DOn't remove!

   \author{Chao Liu
      \inst{1}
   \and Yan Xu
      \inst{1}
   \and Jun-Chen Wan
      \inst{1,2}
   \and Hai-Feng Wang
      \inst{1,2}
   \and Jeffrey L. Carlin
   \inst{3}
   \and Li-Cai Deng
      \inst{1}
   \and Heidi Jo Newberg
      \inst{4}
   \and Zihuang Cao
      \inst{1}
    \and Yonghui Hou
       \inst{5}
    \and Yuefei Wang
       \inst{5}
    \and Yong Zhang
       \inst{5}
   }

   \institute{Key Lab of Optical Astronomy, National Astronomical Observatories, Chinese Academy of Sciences,
             Beijing 100012, China; {\it liuchao@nao.cas.cn}\\
        \and
             University of the Chinese Academy of Sciences, Beijing, 100049, China
        \and
             LSST, 933 North Cherry Avenue, Tucson, AZ 85721, USA
        \and
             Department of Physics, Applied Physics and Astronomy, Rensselaer Polytechnic Institute, Troy, NY 12180, USA
        \and
             Nanjing Institute of Astronomical Optics \& Technology, National Astronomical Observatories, Chinese Academy of Sciences, Nanjing 210042, China
   }

   \date{Received~~2017 month day; accepted~~2017~~month day}

\abstract{We present a statistical method to derive the stellar density profiles of the Milky Way from spectroscopic survey data, taking into account selection effects. We assume that the selection function of the spectroscopic survey is  based on photometric colors and magnitudes and possibly altered during observations and data reductions. Then the underlying selection function for a line-of-sight can be well recovered by comparing the distribution of the spectroscopic stars in a color-magnitude plane with that of the photometric dataset. Subsequently, the stellar density profile along a line-of-sight can be derived from the spectroscopically measured stellar density profile multiplied by the selection function. The method is validated using \emph{Galaxia} mock data with two different selection functions. We demonstrate that the derived stellar density profiles well reconstruct the true ones not only for the full targets, but also for the sub-populations selected from the full dataset. Finally, the method is applied to map the density profiles for the Galactic disk and halo, respectively, using the LAMOST RGB stars. The Galactic disk extends to about $R=19$\,kpc, where the disk still contributes about 10\% to the total stellar surface density. Beyond this radius, the disk smoothly transitions to the halo without any truncation, bending, or broken. Moreover, no over-density corresponding to the Monoceros ring is found in the Galactic anti-center direction. The disk shows moderate north--south asymmetry at radii larger than 12\,kpc. On the other hand, the $R$--$Z$ tomographic map directly shows that the stellar halo is substantially oblate within a Galactocentric radius of 20\,kpc and gradually becomes nearly spherical beyond 30\,kpc.
}

   \authorrunning{Liu et al.}            %author_head in even pages
   \titlerunning{Mapping the Milky Way I}  % title_head in odd pages

   \maketitle
%% The author head (on even pages) and the title head (on odd pages) will be
%% automatically extracted from \author{} and \title{}. Whenever the title is too long,
%% you will be asked to supply a shorter one by inserting either \authorrunning{} or
%% \titlerunning{} before \maketitle. Anyway, you can specify your own heads.
%%
%%
%% Note: In the following text body of your manuscript, please note several differences from
%%       other major journals:
%% (1) \subsection{Please Capitalize the First Letter of Each Notional Word in Subsection Title}
%% (2) Please Capitalize the First Letter of Each Notional Word in all tables' captions

%
%________________________________________________ sections below
%
\section{Introduction}           %% first-level sections will be auto-capitalized
\label{sect:intro}
The stellar density profile is of importance in unveiling the nature of the Galaxy. Either the photometric survey data, which are supposed to be complete to some extent, or the spectroscopic survey data, which can provide more accurate stellar parameters but obviously do not completely cover the sky due to the lower efficiency of the sampling, can be used to count the stars in three-dimensional space.

Although star count modeling has been studied for a long time, not until recently does it give reliable measurements of the Galactic disks and halo based on modern surveys (e.g., Chen et al. \cite{chen2001}, Juri\'c et al. \cite{juric2008}, Watkins et al. \cite{watkins2009}, Bovy et al. \cite{bovy2012}, Xue et al. \cite{xue2015}; also see the review by Bland-Hawthorn \& Gerhard \cite{bland2016}). Juri\'c et al. (\cite{juric2008}) fitted the star counts of various spectral types of stars from the SDSS photometric survey data with two exponential disks and a power-law stellar halo. They found that the scale length of the thin disk is 2.6\,kpc, substantially smaller than that of the thick disk, which is 3.6\,kpc. However, Bovy et al. (\cite{bovy2012}) showed that the scale lengths change from $\sim4.5$ kpc for the mono-abundance populations with small scale heights to $\sim2$\,kpc for those with large scale heights. Comparison between the two results is non-trivial, since the former work identified the thick disk by geometry, while the latter one defined the thick disk based on the chemical abundances. 

It is worthwhile to point out that although the disk profile is often empirically simplified as an exponential or hyperbolic secant form, it is composed of substantially asymmetric structures such as the bar and spiral arms. Furthermore, Lop\'ez-Corredoira et al. (\cite{lopezcorredoira2002}) unveiled that the disk is significantly flared and warped in the outskirts using red clump giant stars selected from the 2MASS catalog (Strutskie et al. \cite{strutskie2006}). Moreover, recent studies also demonstrated vertical asymmetric structures in the outer disk (Xu et al. \cite{xu2015}) as well as in the solar neighborhood (Widrow et al. \cite{widrow2012}).

On the stellar halo, Watkins et al. (\cite{watkins2009}) and Deason et al. (\cite{deason2011}) used RR Lyrae and Blue Horizontal Branch stars, both of which are accurate standard candles, to measure the shape of the stellar halo, respectively. They both found the halo density profile follows a broken power-law with constant axis ratios. Later on, Xue et al. (\cite{xue2015}) suggested that, instead of the broken power-law with constant axis ratio, a single power-law with a radially variable axis ratio can also well fit the density profile of the halo. It is also noted that Xu, Deng \& Hu (\cite{xu2006}, \cite{xu2007}) found that the stellar halo is tri-axial from the star counts using photometric data.

The advantages of using multi-band photometric data in studies of the star counts are that 1) the sampling for the photometric stars is approximately complete within a limiting magnitude; 2) multiple colors can be used to select specific stellar populations and also determine the photometric parallax for the stars; and 3) the large number of the stellar samples can improve the statistical performance. However, there are a few disadvantages of photometric data, including: 1) it is hard to accurately select stellar populations based on chemical abundances or ages; 2) the dwarf/giant separation is difficult in most broad-band photometric data; 3) the distance estimation based on photometric data in principle suffers from larger uncertainty and potential systematic bias; and 4) binarity may affect the results of the star counts since unresolved binaries usually show slightly different colors than single stars. %Therefore, the study of the stellar density profile using spectroscopic survey data is complimentary with those based on the photometric data.

In general, spectroscopic data can provide relatively precise stellar parameter estimates, which lead to better distance determinations. Moreover, the stellar parameters derived from spectroscopic data are helpful in better identifying a stellar population. Therefore, star counts based on spectroscopic data should provide a valuable complement to photometric studies. It can play even more important role in understanding the evolution of the Galaxy via the structures of different stellar populations. However, because it is difficult to sample large regions of sky spectroscopically to a completeness that can easily be achieved by a photometric survey, the stellar density measurements from a spectroscopic survey highly depend on correction of the selection function explicitly or implicitly induced in the survey. 

Techniques to correct for the selection function have been discussed in many works. Liu \& van de Ven (\cite{liu2012}) corrected the selection effects for SEGUE G-dwarf stars by comparing the stellar density in color-magnitude-distance grids between the spectroscopic and photometric survey data. A similar method has been used in Zhang et al. (\cite{zhang2013}) for the measurement of the local dark matter density. Bovy et al. (\cite{bovy2012}) proposed a Bayesian technique to model the stellar density profiles with analytic forms accounting for the selection effects. This technique was later applied to the APOGEE survey data (Bovy et al. \cite{bovy2016}) and SEGUE K-giant stars (Xue et al. \cite{xue2015}). It is noted that Xia et al. (\cite{xia2016}) also used a similar technique with the LAMOST survey data.

Recently, the LAMOST survey (Cui et al. \cite{cui2012}, Zhao et al. \cite{zhao2012}, Deng et al. \cite{deng2012}) has collected more than 5 million low resolution stellar spectra in the DR3 catalog. These stars cover much of the stellar halo to Galactocentric radius of 80--100\,kpc (Liu et al. \cite{liu2014}) and cover the Galactic anti-center region to as far as 50\,kpc. This provides a remarkably large number of distant stars with which to measure the shapes of the outer disk and halo of the Milky Way. In this work, we revise the approach of stellar density measurement from spectroscopic survey data originally proposed by Liu \& van de Ven (\cite{liu2012}) and re-organize it in terms of Bayesian statistics. The post-observation selection function is taken into account during the determination of the stellar density. Meanwhile, we do not presume any analytic forms for the density profiles of neither the disks nor the halo. This allows us to better detect the potentially more complicated shapes of the structures, e.g., asymmetric features in the disk or the possibly radial variation of the axis ratio of the halo, in a non-parametric manner.

The paper is organized as follows. In Section~\ref{sect:method}, we propose the method to derive the stellar density profile along a given line-of-sight with the consideration of the selection function. In Section~\ref{sect:valid}, we validate the method using mock data with various selection functions. In Section~\ref{sect:density}, we apply it to the LAMOST red giant branch stars and demonstrate the derived stellar density map for the disk and halo components. In section~\ref{sect:discussion}, we discuss the smoothing effect of the LAMOST plate due to its large field-of-view and how to deal with the selection function slightly tuned by the observations. Finally, we draw brief conclusions in the last section.

\section{Stellar density profile along a line-of-sight}\label{sect:method}

Given a line-of-sight with Galactic coordinates ($l$, $b$), we assume that the  photometric data is always complete within the limiting magnitude\footnote{The limiting magnitude of a photometric survey is usually defined by the magnitude at signal-to-noise of 5 or 10\ $\sigma$.}. We also assume that the selection of the spectroscopic targets  is only associated with the color-magnitude diagram.  The selection function as a function of the color(s) and magnitude(s)  could be complicated. In principle, observations and data reduction may lose some data and thus the final selection function is slightly tuned. However, such a change would not induce substantial selection bias in stellar metallicity, age or kinematics, which are critical for distinguishing various stellar populations. Therefore the color-magnitude based selection function slightly tuned by the observations and data processing would not induce systematics in stellar populations. It is convenient to assume that it is a continuous function so that it is approximately constant in a sufficiently small region around color index $c$ and magnitude $m$.

\subsection{Generality}
The probability density distribution (PDF) $p_{ph}(D|c,m,l,b)$ represents the probability to find a star at the distance $D$ at given $c$, $m$,  $l$, and $b$. The probability to find a star in a small volume ($D$, $D+\Delta D$) is then written as
\begin{eqnarray}\label{eq:Pr}
Pr([D,D+\Delta D]|c,m,l,b)=&\int_{D}^{D+\Delta D}{p_{ph}(D|c,m,l,b)dD}\nonumber\\
\doteq&p_{ph}(D|c,m,l,b)\Delta D.
\end{eqnarray}
On the other hand, $Pr([D, D+\Delta D]|c,m,l,b)$ is also equivalent to the fraction of the stars, which can be calculated from the underlying stellar density profile, located within the small volume. Thus, it can also be written as  
\begin{equation}\label{eq:pph0}
Pr([D,D+\Delta D]|c,m,l,b)={\nu_{ph}(D|c,m,l,b)\Omega D^2\Delta D\over{\int_{0}^{\infty}{\nu_{ph}(D|c,m,l,b)}\Omega D^2dD}}.
\end{equation}
where $\nu_{ph}$ is the volume stellar density measured from the photometric stars (i.e., the complete dataset), $\Omega D^2\Delta D$ is the volume element between $D$ and $D+\Delta D$, and $\Omega$ is the solid angle of the line-of-sight. Combining Eqs.~(\ref{eq:Pr}) with~(\ref{eq:pph0}), we have
\begin{equation}\label{eq:pph}
p_{ph}(D|c,m,l,b)\Delta D={\nu_{ph}(D|c,m,l,b)\Omega D^2\Delta D\over{\int_{0}^{\infty}{\nu_{ph}(D|c,m,l,b)}\Omega D^2dD}}.
\end{equation}
Similarly, for the spectroscopic data, Eq.~(\ref{eq:pph}) becomes
\begin{equation}\label{eq:psp}
p_{sp}(D|c,m,l,b)\Delta D={\nu_{sp}(D|c,m,l,b)\Omega D^2\Delta D\over{\int_{0}^{\infty}{\nu_{sp}(D|c,m,l,b)}\Omega D^2dD}},
\end{equation}
where $\nu_{sp}$ represents the stellar density measured from counting the spectroscopic survey stars only. Within the small region around $c$ and $m$, the selection function is assumed to be flat with either the color index or magnitude. Thus, it does not change the probability to find a star in both the photometric and spectroscopic  samples, i.e.,
\begin{equation}\label{eq:pp}
p_{ph}(D|c,m,l,b)=p_{sp}(D|c,m,l,b).
\end{equation}
Then, Eqs.~(\ref{eq:pph}) and~(\ref{eq:psp}) can be combined together via Eq.~(\ref{eq:pp}):
\begin{equation}\label{eq:nuPDF}
\nu_{ph}(D|c,m,l,b)=\nu_{sp}(D|c,m,l,b)S^{-1}(c,m,l,b),
\end{equation}
where 
\begin{equation}\label{eq:Sel}
S(c,m,l,b)={{\int_{0}^{\infty}{\nu_{sp}(D|c,m,l,b)}\Omega D^2dD}\over{\int_{0}^{\infty}{\nu_{ph}(D|c,m,l,b)}\Omega D^2dD}}
\end{equation}
is the selection function at $c$ and $m$ along ($l$, $b$).% It is noted that if bringing Eq.~(\ref{eq:nuPDF}) to~(\ref{eq:pph}), then we can obtain the equivalent form with Equation (4) of Bovy et al. (\cite{bovy2012}). 

Then the stellar density profile for all photometric stars along ($l$, $b$) can be derived by integrating over $c$ and $m$:
\begin{equation}\label{eq:nuall}
\nu_{ph}(D|l,b)=\iint{\nu_{sp}(D|c,m,l,b)S^{-1}(c,m,l,b)dcdm}.
\end{equation}

\subsection{Consideration of a stellar population}
Now consider a stellar sub-population $C$ selected from spectroscopic data based on, for example, the metallicity, luminosity, or age. Following Bayes' theorem, the probability to find a star at $D$ given $C$, $c$, $m$, $l$, and $b$ is written as
\begin{equation}\label{eq:pphC}
p_{ph}(D|C,c,m,l,b)={p_{ph}(C|D,c,m,l,b)p_{ph}(D|c,m,l,b)\over{p_{ph}(C|c,m,l,b)}}.
\end{equation}
If there is no special selection function for $C$ members at given $c$ and $m$ along ($l$, $b$), then we would expect that $p_{ph}(C|D, c, m, l, b)$, i.e. the probability about $C$ at given $D$, $c$, $m$, $l$, and $b$, should be same as that for the spectroscopic data, $p_{sp}(C|D,c,m,l,b)$. Subsequently, by integrating over $D$, we obtain that $p_{ph}(C|c,m,l,b)=p_{sp}(C|c,m,l,b)$. Finally, applying Eq.~(\ref{eq:pp}) to this equation, all the terms of $p_{ph}$ in the right hand side of Eq.~(\ref{eq:pphC}) can be replaced  with $p_{sp}$. Applying Bayes' theorem again to the right hand side, we obtain
\begin{equation}\label{eq:pC}
p_{ph}(D|C,c,m,l,b)=p_{sp}(D|C,c,m,l,b).
\end{equation}
Therefore, similar to Eq.~(\ref{eq:nuPDF}), we infer that
\begin{equation}\label{eq:nuPDFC}
\nu_{ph}(D|C,c,m,l,b)=\nu_{sp}(D|C,c,m,l,b)S^{-1}(C,c,m,l,b),
\end{equation}
where
\begin{equation}\label{eq:CS}
S(C,c,m,l,b)={{\int_{0}^{\infty}{\nu_{sp}(D|C,c,m,l,b)}\Omega D^2dD}\over{\int_{0}^{\infty}{\nu_{ph}(D|C,c,m,l,b)}\Omega D^2dD}}.
\end{equation}	
And consequently,
\begin{equation}\label{eq:nuallC}
\nu_{ph}(D|C,l,b)=\iint{\nu_{sp}(D|C,c,m,l,b)S^{-1}(C,c,m,l,b)dcdm}.
\end{equation}
%Notice that the selection function $S(c,m,l,b)$ is independent of the selection of the sub-population $C$.

\subsection{Estimation of $\nu_{sp}$}
In Eqs~(\ref{eq:nuall}) and~(\ref{eq:nuallC}), $\nu_{ph}$, as the density profile for the complete dataset, is the one to be determined. We can estimate it by measuring $\nu_{sp}$ from the spectroscopic data. Usually, along a given line-of-sight, spectroscopic observations can only target a few hundreds to a few thousands of stars. This implies that $\nu_{sp}$ has to be established from a relatively small dataset. In order to deal with such a situation, we derive $\nu_{sp}$ by using the kernel density estimation (KDE). The KDE method can account for both the uncertainties in the distance and the relatively small number of stars in a given small region around ($c$, $m$). At ($l$, $b$), the PDF of $D$ is contributed by all the stars around $c$ and $m$ in terms of
\begin{equation}\label{eq:KDE}
p_{sp}(D|c,m,l,b)={1\over{n_{sp}(c,m,l,b)}}\sum_{i}^{n_{sp}(c,m,l,b)}{p_i(D)},
\end{equation}
where $p_i(D)$ is the PDF of the distance estimate for the $i$th star and $n_{sp}(c,m,l,b)$ is the number of the spectroscopic stars at a given $c$, $m$, $l$, and $b$.
Bringing Eq.~(\ref{eq:KDE}) into Eq.~(\ref{eq:psp}), we obtain
\begin{equation}\label{eq:KDE2}
\nu_{sp}(D|c,m,l,b)={1\over{\Omega D^2}}\sum_{i}^{n_{sp}(c,m,l,b)}{p_i(D)}.
\end{equation}
Note that $\int{\nu_{sp}(c,m,l,b)\Omega D^2dD}=n_{sp}(c,m,l,b)$, they are then cancelled in Eq.~(\ref{eq:KDE2}).

In principle, the KDE determined $\nu$ is a continuous function and thus can read value at any $D$. However, if there are very few stars nearby $D$ or the given $D$ is beyond the farthest star, the derived $\nu$ is not well constrained by observations and thus may not be reliable any more. Therefore, in practice, the stellar density value is sensible only at the positions with sufficient star samples. For convenience, we only adopt $\nu$ at the  distance in which the spectroscopic stars are  located.

It is worthy to note that the stellar density derived for a given ($l$, $b$) is averaged over the solid angle of the field-of-view, $\Omega$. Therefore, the derived density at the distance in which a star is located does not equal to the stellar density for the exact 3D spatial position of the star, but only an approximation of the latter.

\subsection{Estimation of $S$}\label{sect:S}
Theoretically, $S$ is defined by Eqs.~(\ref{eq:Sel}) and~(\ref{eq:CS}). Practically, it is not possible to calculate $S$ directly from $\nu_{ph}$ and $\nu_{sp}$. Notice that the integration of $\nu$ is actually equivalent with the total number of stars at given $c$, $m$, $l$, $b$. Therefore, $S$ can be evaluated from
\begin{equation}\label{eq:S2}
S(c,m,l,b)={n_{sp}(c,m,l,b)\over{n_{ph}(c,m,l,b)}}.
\end{equation}
For the case of sub-population $C$, $p_{ph}(C|c,m,l,b)=p_{sp}(C|c,m,l,b)$, we then have
\begin{equation}\label{eq:SC2}
{n_{ph}(C,c,m,l,b)\over{n_{ph}(c,m,l,b)}}={n_{sp}(C,c,m,l,b)\over{n_{sp}(c,m,l,b)}}.
\end{equation}
Hence, we infer that $S(C,c,m,l,b)=S(c,m,l,b)$.

In order to improve the precision of the computation in Eqs.~(\ref{eq:nuall}) and (\ref{eq:nuallC}), we calculate $S^{-1}$ instead of $S$ in the rest of the paper.

\section{Validations with \emph{Galaxia} mock data}\label{sect:valid}
Before applying it to the real data, we validate the method using mock star catalog. We select a 20-square degree area toward the north Galactic pole from \emph{Galaxia} simulation (Sharma et al. \cite{sharma2011}) and obtain 10\,987 mock stars with $K$ magnitude brighter than 15\,mag. The \emph{Galaxia} mock catalog contains the distances, stellar parameters, and ages, which can be used to test whether the derived density profiles are precise for some specially selected stellar populations.

We test the method with two different selection functions, which are discussed separately in the following subsections.

\subsection{Tests with the selection function \textbf{T1}}\label{sect:T1}
\begin{figure}[htbp]
\centering
\includegraphics[scale=0.5]{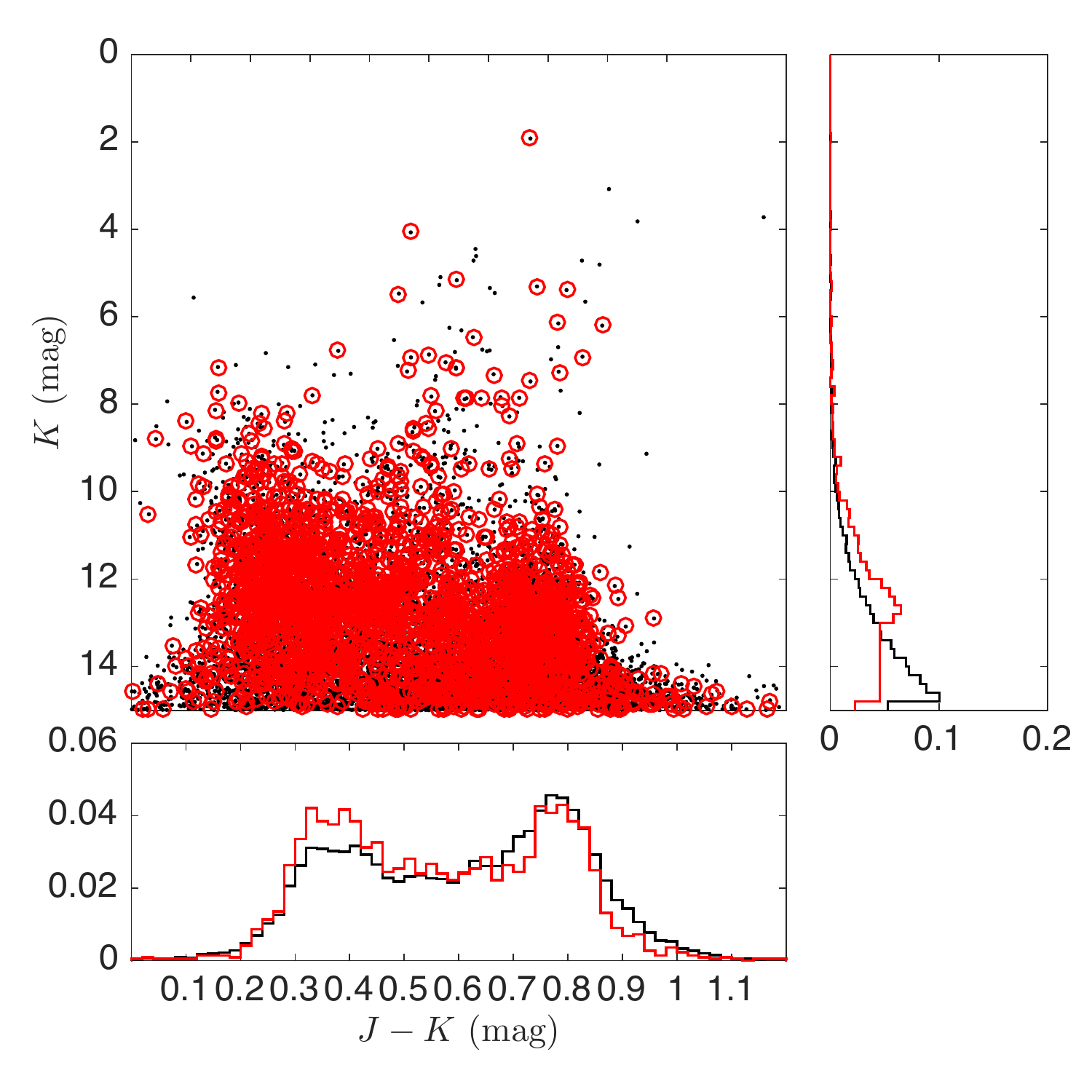}
\caption{The black dots show the distribution in $J-K$ vs.$K$ for the complete sample of mock stars generated from \emph{Galaxia}. The red circles indicate the mock spectroscopic stars according to the selection function \textbf{T1}. The right panel shows the normalized distributions of $K$ for the complete and selected mock stars with black and red lines, respectively. The bottom panel indicates the normalized distributions of $J-K$ for the two populations with same colors as in the right panel.}\label{fig:cmd1}
\end{figure}

\begin{figure*}[htbp]
\centering
\includegraphics[scale=0.4]{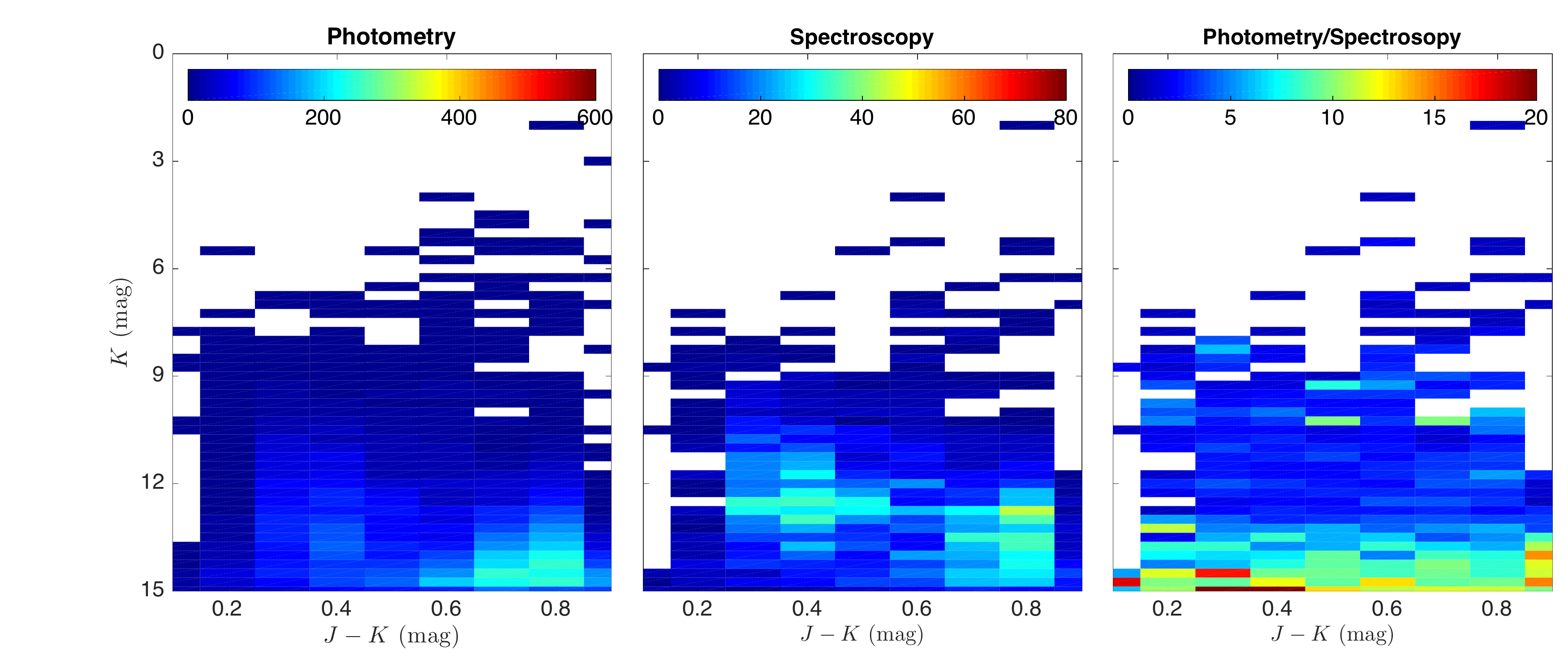}
\caption{The left panel shows the number density of stars in $J-K$ vs. $K$ diagram for the complete mock stars. The color codes the number of stars within each small $J-K$ vs. $K$ bin. The middle panel shows the similar map of the number density in the $J-K$ vs. $K$ plane for the mock spectroscopic stars based on the selection function \textbf{T1}. The right panel displays the map of $S^{-1}(J-K, K)$ for \textbf{T1}. The color scales are indicated in the colorbars located above the panel.}\label{fig:cmdnumbers1}
\end{figure*}

\begin{figure*}[htbp]
\centering
\includegraphics[scale=0.5]{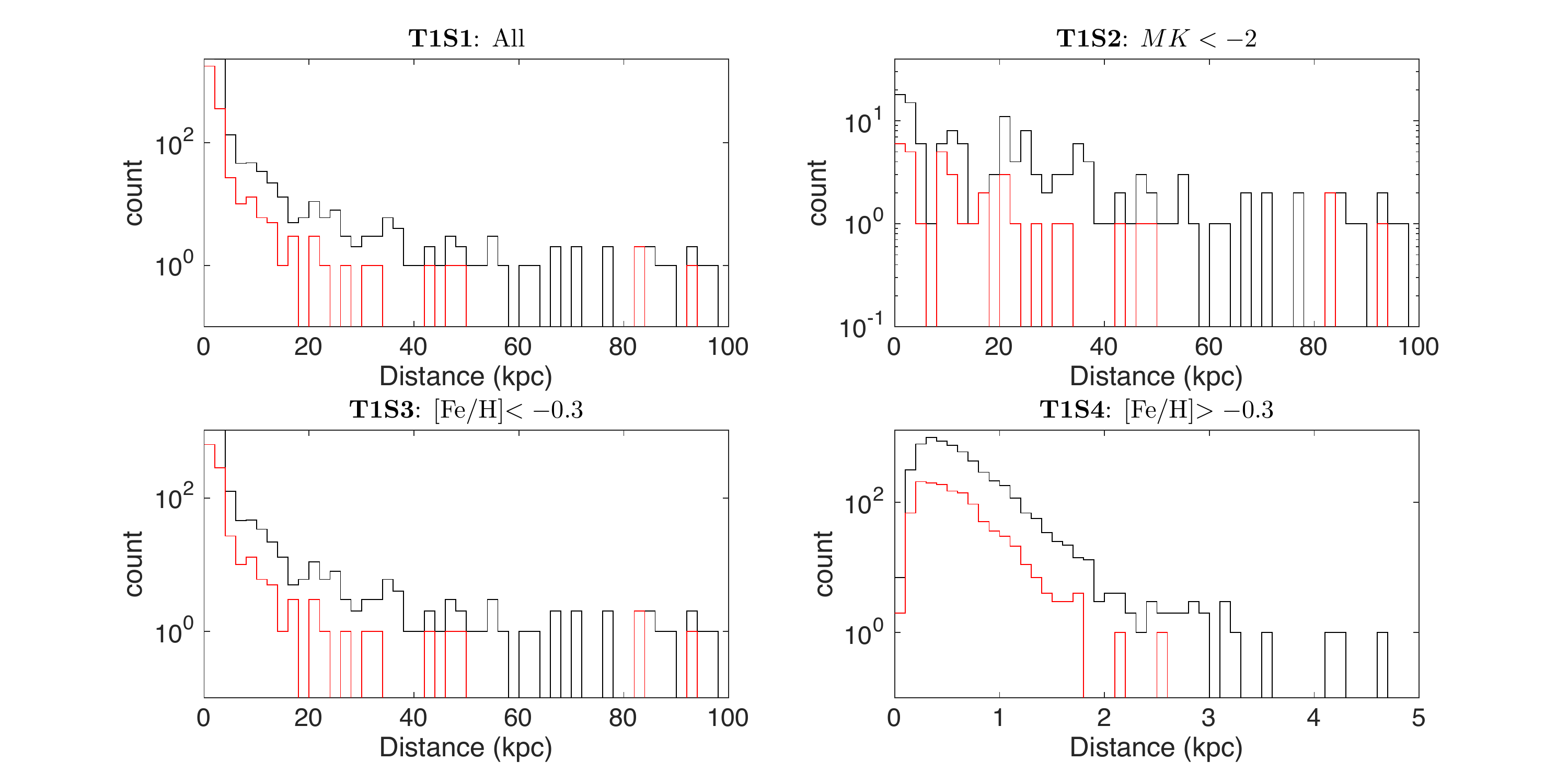}
\caption{The 4 panels indicate the histogram of $D$ for the mock spectroscopic (red lines) and complete (black lines) dataset, respectively. The top-left, top-right, bottom-left, and bottom-right panels displays the histograms for the populations \textbf{T1S1}, \textbf{T1S2}, \textbf{T1S3}, and \textbf{T1S4}, respectively.}\label{fig:distdistribution1}
\end{figure*}

\begin{figure*}[htbp]
\centering
\begin{minipage}{18cm}
\centering
\includegraphics[scale=0.4]{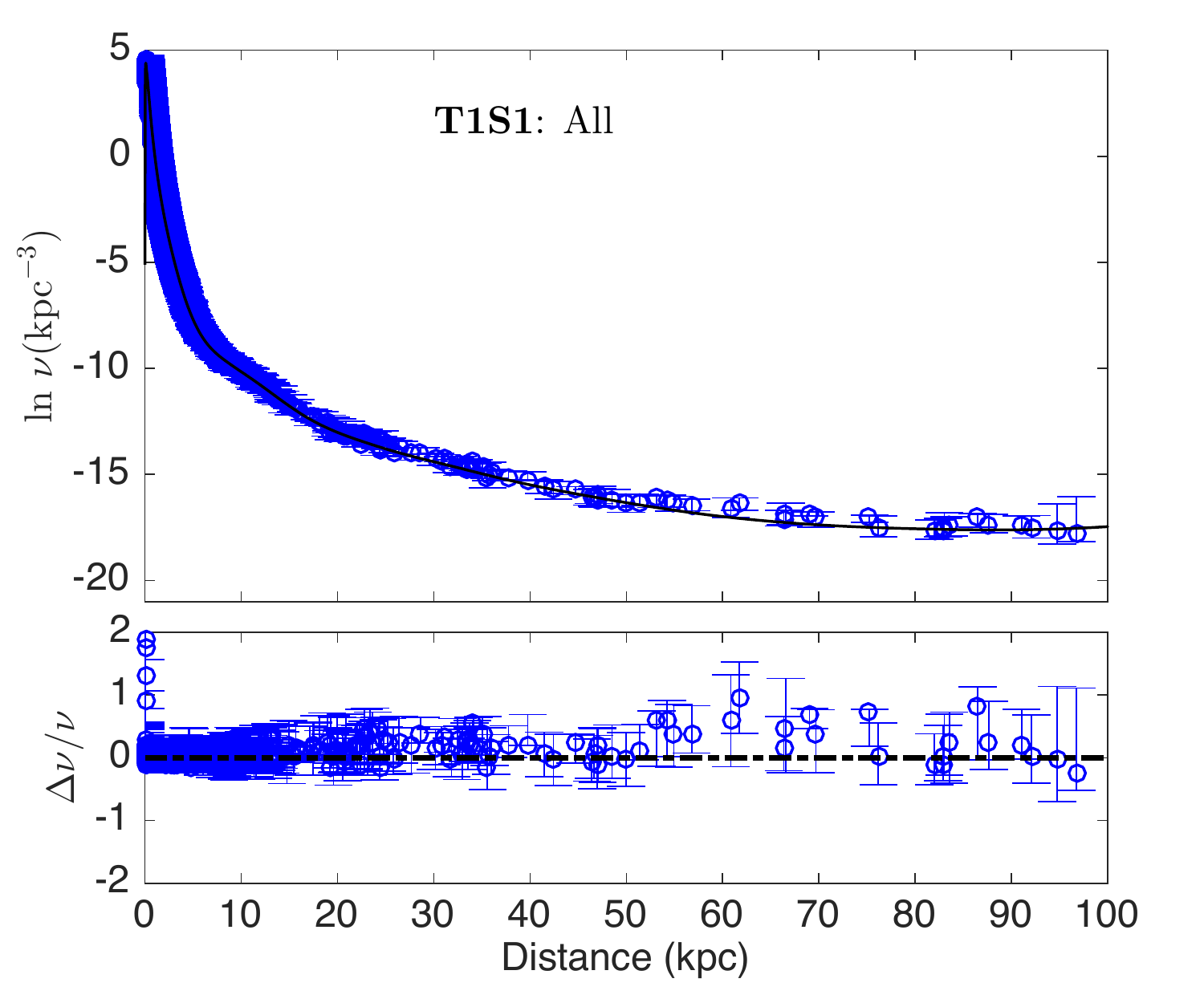}
\includegraphics[scale=0.4]{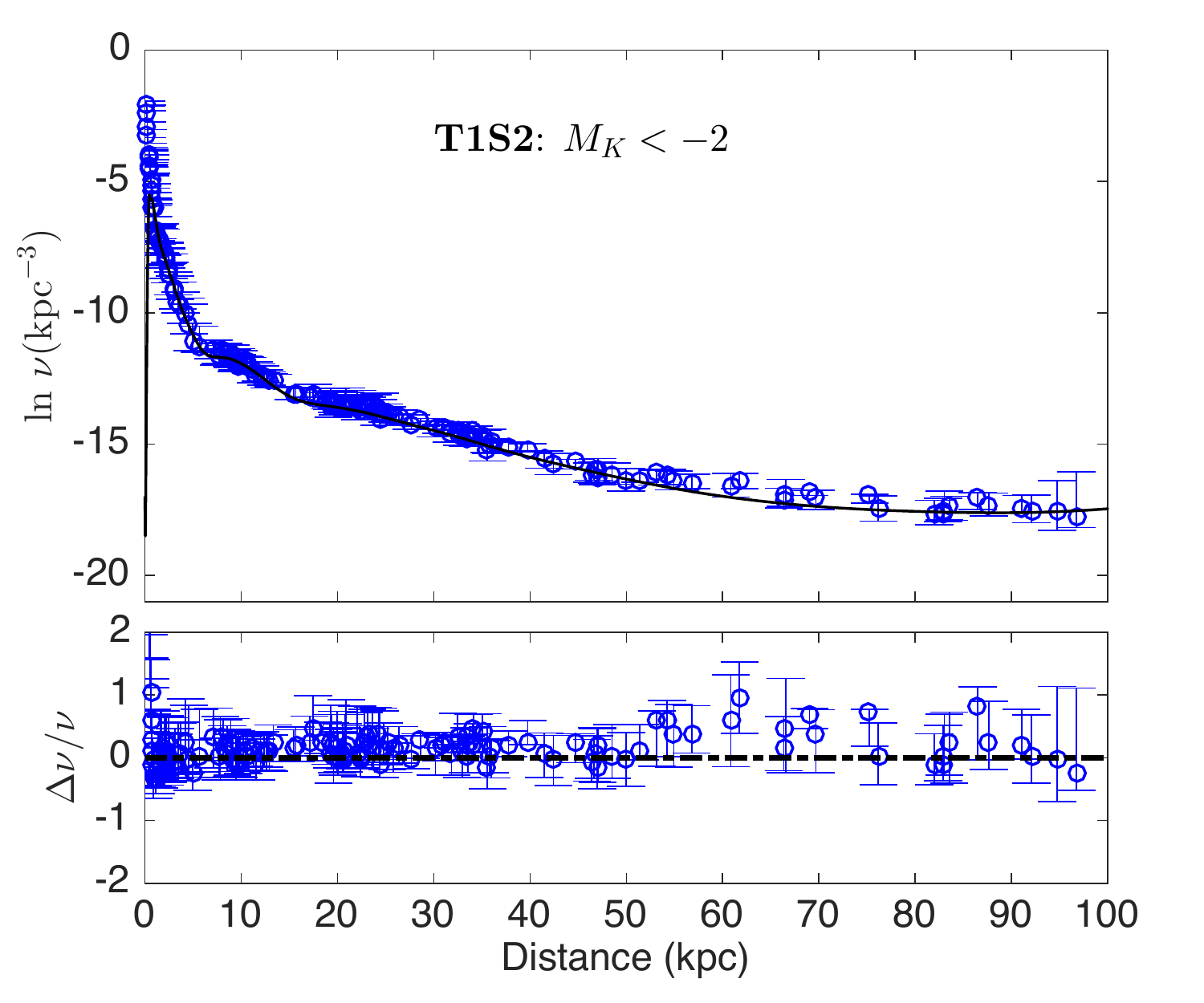}
\includegraphics[scale=0.4]{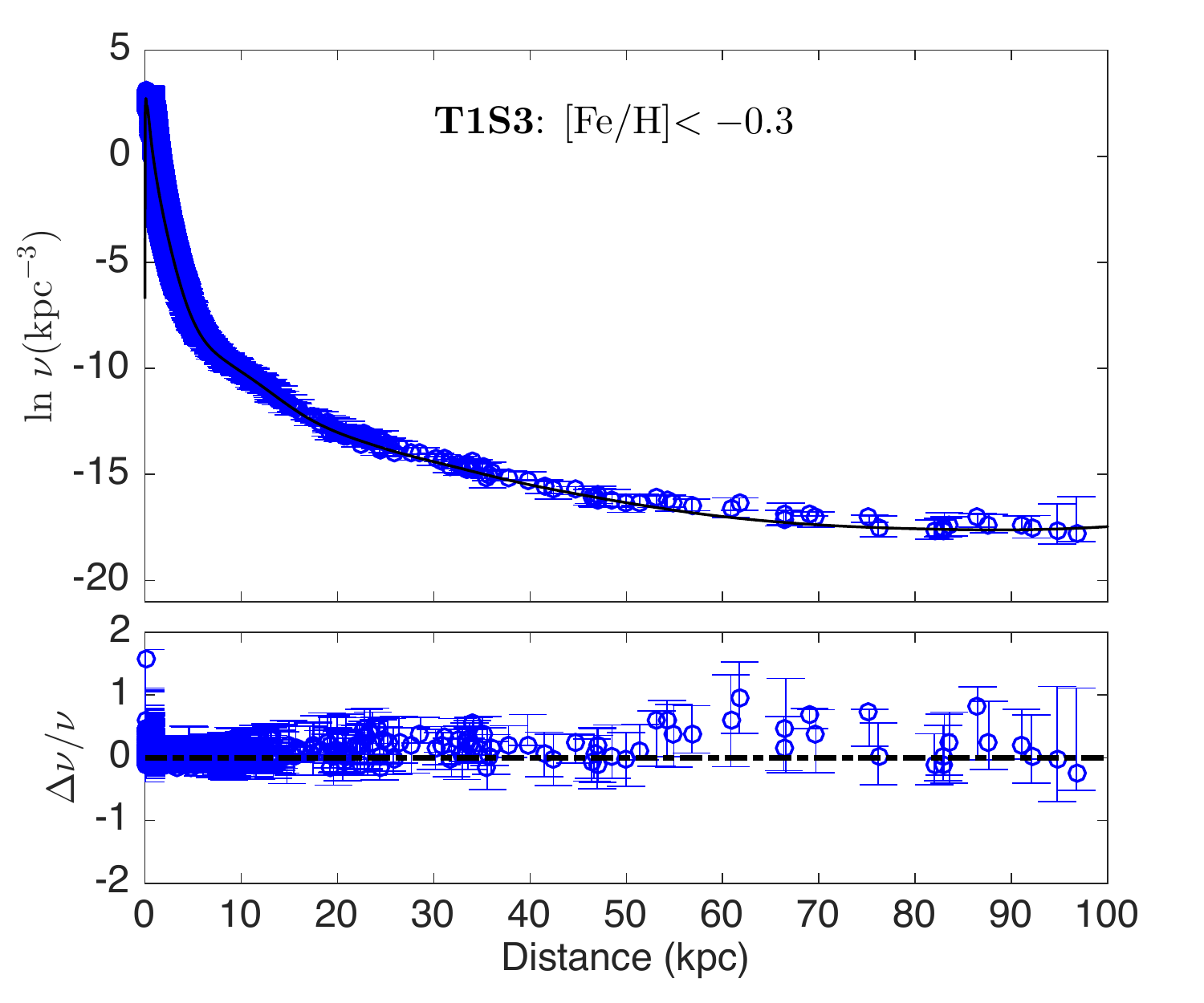}
\includegraphics[scale=0.4]{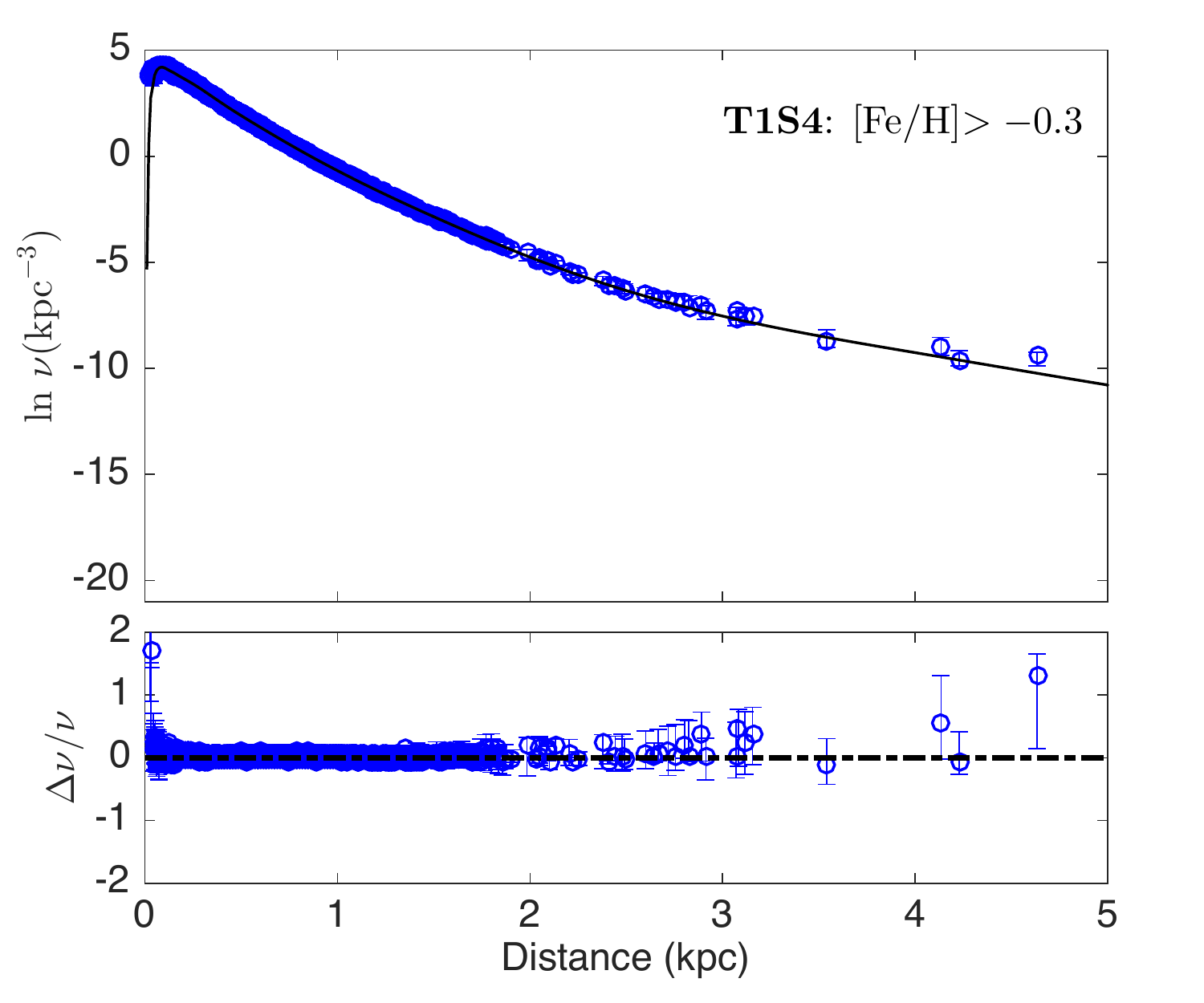}
\end{minipage}
\caption{The 4 panels show the derived density profiles along distances. The top-left, top-right, bottom-left, and bottom-right panels display the profiles derived from the mock spectroscopic stars (blue circles with error bars) and the ``true'' profile from the complete dataset (black lines) for the populations \textbf{T1S1}, \textbf{T1S2}, \textbf{T1S3}, and \textbf{T1S4}, respectively. The relative residual density, $\Delta\nu/\nu$, for each case is displayed in the bottom of the corresponding panel. 
%The blue circles with error bars indicate the derived stellar densities right at the spatial positions in which the mock spectroscopic stars are located.
	}\label{fig:nu1}
\end{figure*}

Before the tests, we construct the selection function to mimic the real spectroscopic survey. The selection of the mock spectroscopic stars is based on the infrared band $J$ and $K$. First, we do not induce any selection in the color index $J-K$. Second, the selection function for $K$ is divided into two parts at $K=13$\,mag. At $K<13$\,mag, we arbitrarily select stars, while at $K>13$\,mag,  we apply a flat selection function to keep the mock spectroscopic stars evenly distributed along $K$. This mimics the strategy that the targeting is more bias to the bright rather than to the faint sources. We denote such selection function as \textbf{T1}. 

We randomly draw 50 groups of the mock spectroscopic samples according to \textbf{T1} and derive the density profile for each group of samples. As a sample, the mock spectroscopic stars from one of the 50 groups are highlighted with red circles in Fig.~\ref{fig:cmd1}, in which the black dots indicate the $J-K$ vs. $K$ distribution for the complete dataset. The right panel shows the normalized distributions of $K$ for the mock spectroscopic stars (red line) and the complete dataset (black line), respectively. It can be seen that the red line shows a similar trend as the black one at $K<13$ and then become flat at $K>13$. This reflects the definition of the selection function \textbf{T1}. The normalized distributions of $J-K$ in the bottom panel display that even though we do not apply any selection function to the color index, the normalized distribution of $J-K$ for the mock spectroscopic stars (red line) is not exactly same as that for the full mock dataset (black line). This is a natural result of the selection function in $K$. There are more red stars in the regime of $K>13$ than at $K<13$. The selection strategy for $K>13$ reduces the sampling rate of the fainter and redder stars in the mock spectroscopic stars. This leads to a slightly different distribution in $J-K$ with the complete dataset. 

Fig.~\ref{fig:cmdnumbers1} shows how the selection function $S$ is calculated for the same group of the mock spectroscopic stars as in Fig.~\ref{fig:cmd1}. In the left panel, the color shows the number density of stars for the complete sample, i.e., $n_{ph}(J-K,K,l,b)$. The size of each bin is $\Delta(J-K)=0.1$ and $\Delta K=0.25$. It shows a high density at $J-K\sim0.8$ and $K\sim14$, which is most likely contributed by K/M dwarf stars. Applying the selection function \textbf{T1}, the map of the number density of mock spectroscopic stars, $n_{sp}(J-K,K,l,b)$, is shown in the middle panel. The right panel shows the map of $S^{-1}(J-K,K,l,b)$, which is obtained by dividing the left by the middle panel of Fig.~\ref{fig:cmdnumbers1}. It is seen that $S^{-1}(J-K,K,l,b)$ is roughly separated into two platforms, one is at $K<13$ with the value of about 5 and the other is at $K>13$ with the value of about 10. This pattern implies that the selection function \textbf{T1} does not induce selection effects in $J-K$ at each $K$ bin, although the overall distributions in $J-K$ (as shown in the bottom panel of Fig.~\ref{fig:cmd1}) are slightly different than the complete dataset.

In order to mimic a more realistic situation, i.e., study the stellar density via a selected stellar population, we select 4 different populations defined in Table~\ref{tab:sample} for tests. Table~\ref{tab:sample} also lists the median numbers and scatter of the 50 groups of the arbitrarily selected mock spectroscopic samples based on \textbf{T1} in the 3rd column. It is noted that for the sub-population \textbf{T1S2}, which is selected with $M_K<-2$, only about 40 stars are involved in the determination of the stellar density profile. This is a good sample to test the performance of the density estimates for a small dataset.

\begin{table}
\centering\caption{The test populations selected from the \emph{Galaxia} mock data.}\label{tab:sample}
\begin{tabular}{c|c|c|c}
\hline\hline
Population & Criteria & Number for \textbf{T1} & Number for \textbf{T2}\\
\hline
\textbf{S1} & all spectral types & 2230$\pm29$ & 1943$\pm22$\\
\textbf{S2} & $M_K<-2$ & 40$\pm5$ & 23$\pm3$\\
\textbf{S3} & [Fe/H]$<-0.3$ & 979$\pm23$ & 998$\pm22$ \\
\textbf{S4} & [Fe/H]$>-0.3$ & 1244$\pm29$ & 946$\pm17$\\
\hline\hline
\end{tabular}
\end{table}

Fig.~\ref{fig:distdistribution1} shows the distributions of star counts without a selection function correction for the complete set (black lines) and for one of the 50 groups of the mock spectroscopic samples (red lines), respectively. The 4 panels show the cases for the 4 different sub-populations, respectively. It displays that the selection function \textbf{T1} does distort the spatial distributions of the mock spectroscopic stars from that of the complete samples.

The panels in Fig.~\ref{fig:nu1} show the derived median stellar densities and their 1$\sigma$ dispersions (blue circles with error bars) over the 50 groups of the mock spectroscopic stars at the spatial positions in which the mock spectroscopic stars are located. In contrast, the corresponding stellar densities for the complete dataset, i.e., the ``true'' profiles, are superposed as black solid lines in the panels. In the bottom of each panel, the  relative $\Delta\nu/\nu$ is also displayed. It is seen that the derived stellar density profiles from the mock spectroscopic data are perfectly consistent with the ``true'' profiles. This confirms that when the selection function is very simple, the stellar density can be well reconstructed. In particular for the case \textbf{T1S2}, the accuracy of the derived stellar density is still very high even though the sample contains only about 40 stars (see the top-right panel of Fig.~\ref{fig:nu1}). %For the case \textbf{T1}\textbf{S2}, because the less number of stars are included in the population, larger sampling noise leads to systematic bias to some level. The internal uncertainty of the method is very small. In fact, the error bars, which are too small to be seen, are overlapped with the blue circles. 

\subsection{Tests with selection function \textbf{T2}}\label{sect:T2}
\begin{figure*}
\centering
\includegraphics[scale=0.45]{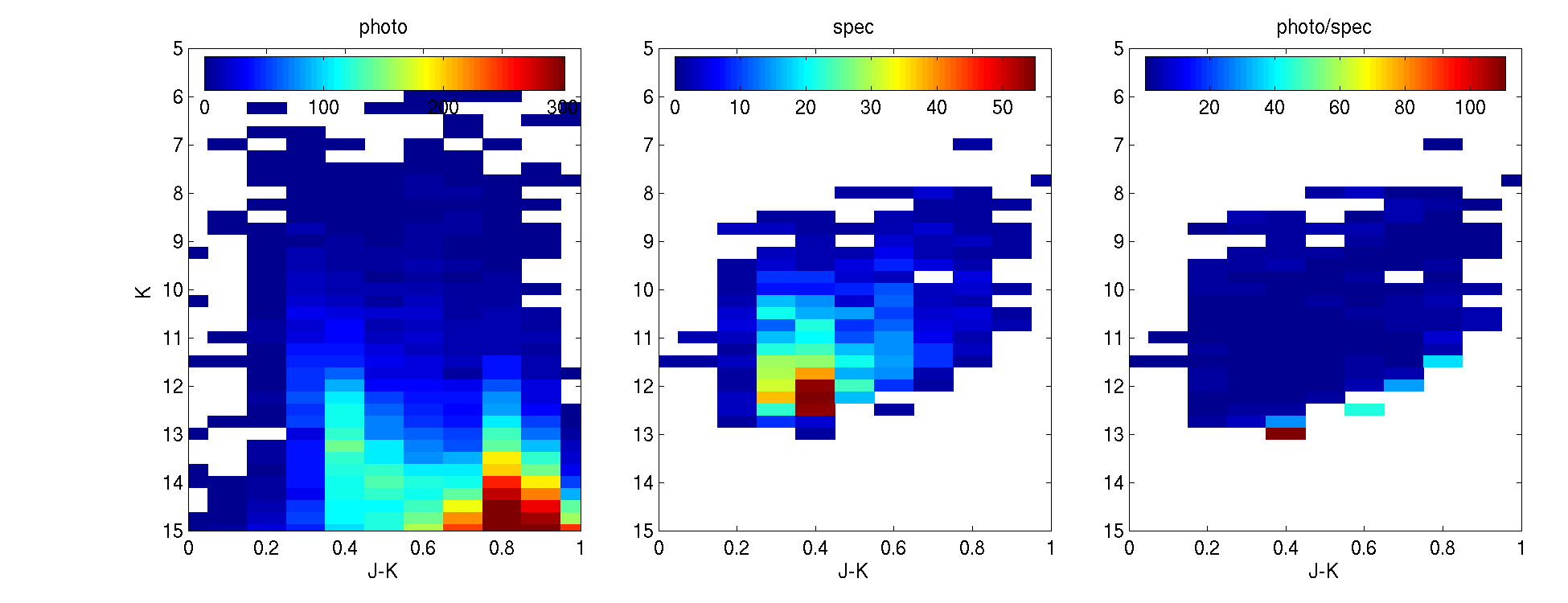}
\caption{A sample of the selection function of a LAMOST plate ``HD103843N315834V01''. The left panel is the $J-K$ vs. $K$ diagram for the photometric data in the same line-of-sight. The middle panel displays the spectroscopically selected data. The right panel shows $S^{-1}$.}\label{fig:LMS}
\end{figure*}
\begin{figure}[htbp]
\centering
\includegraphics[scale=0.5]{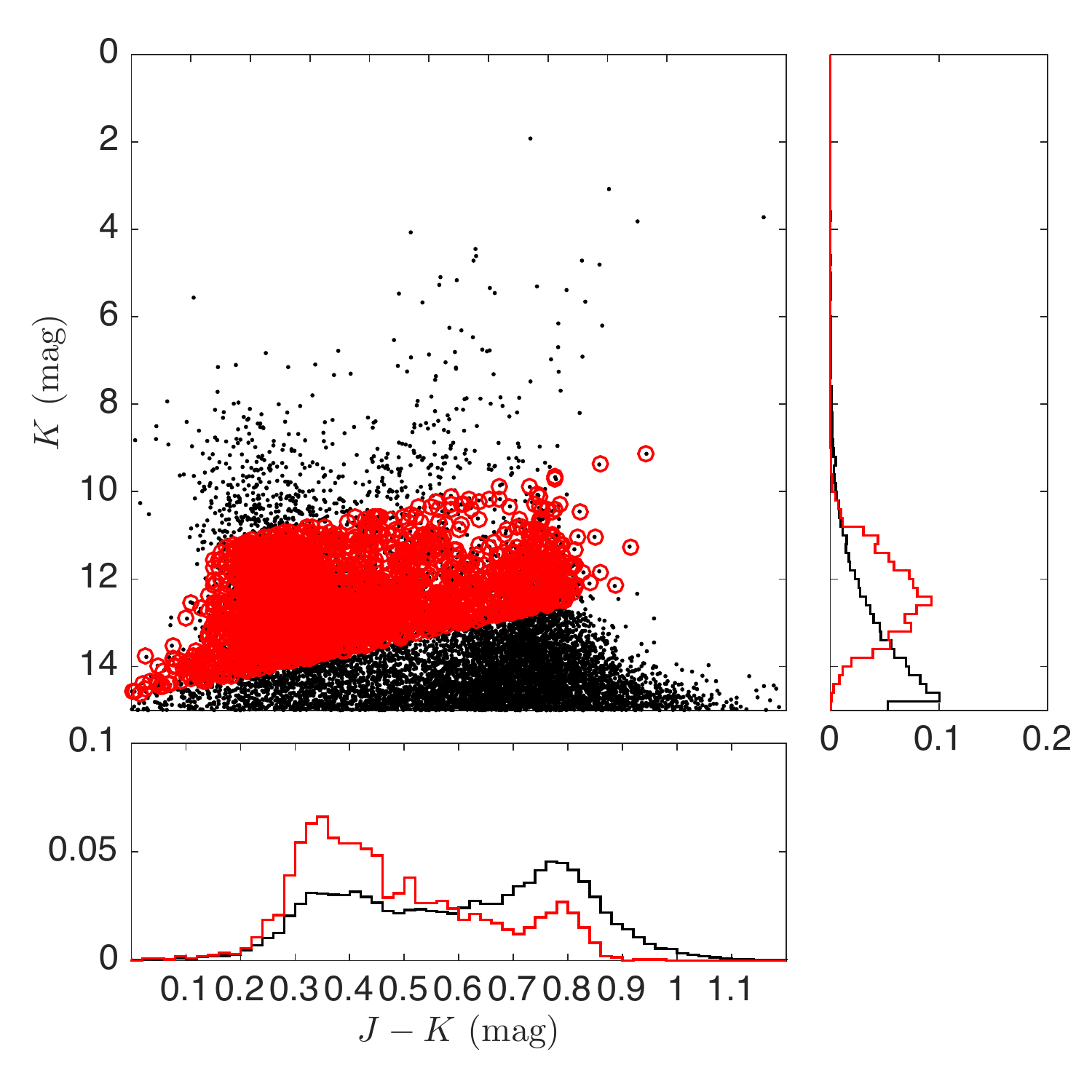}
\caption{Similar to Fig.~\ref{fig:cmd1}, but with the selection function \textbf{T2}.}\label{fig:cmd2}
\end{figure}

\begin{figure*}[htbp]
\centering
\includegraphics[scale=0.4]{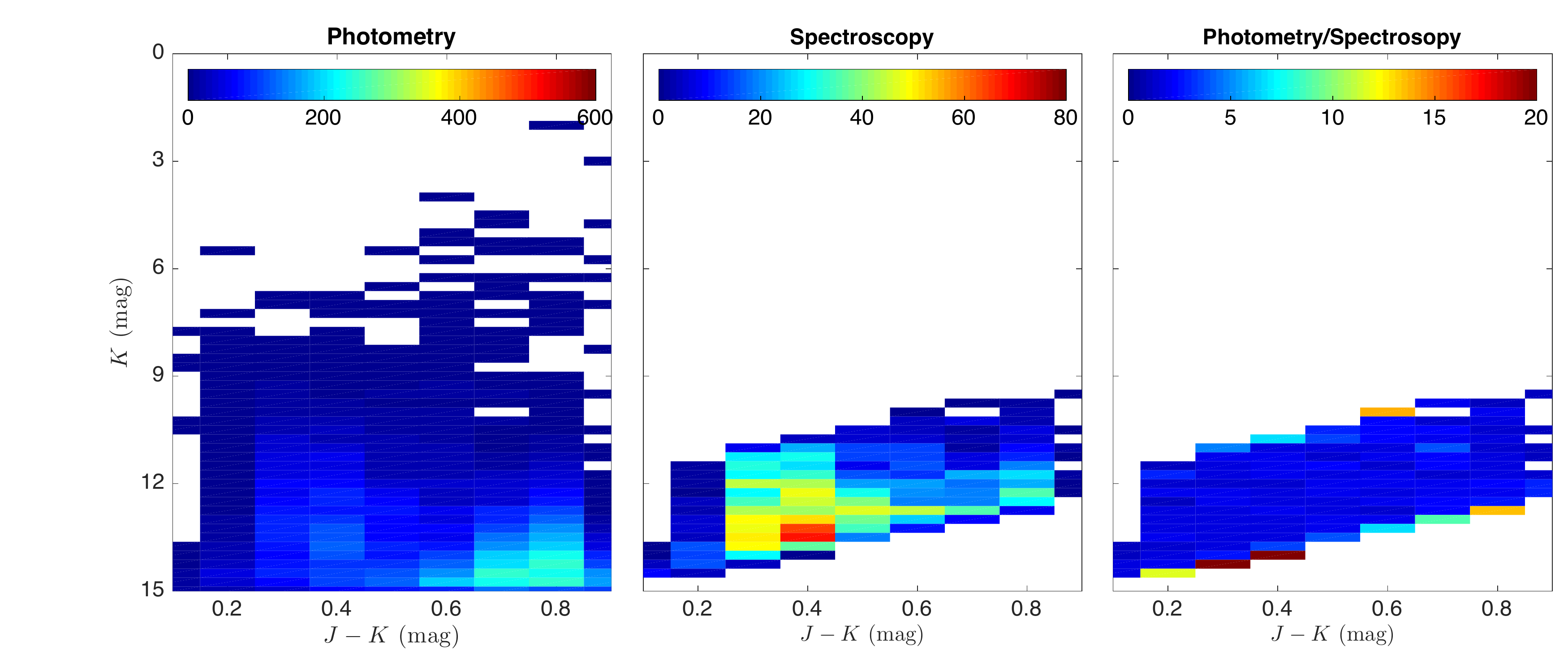}
\caption{Similar to Fig.~\ref{fig:cmdnumbers1}, but with the selection function \textbf{T2}.}\label{fig:cmdnumbers2}
\end{figure*}

\begin{figure*}[htbp]
\centering
\includegraphics[scale=0.5]{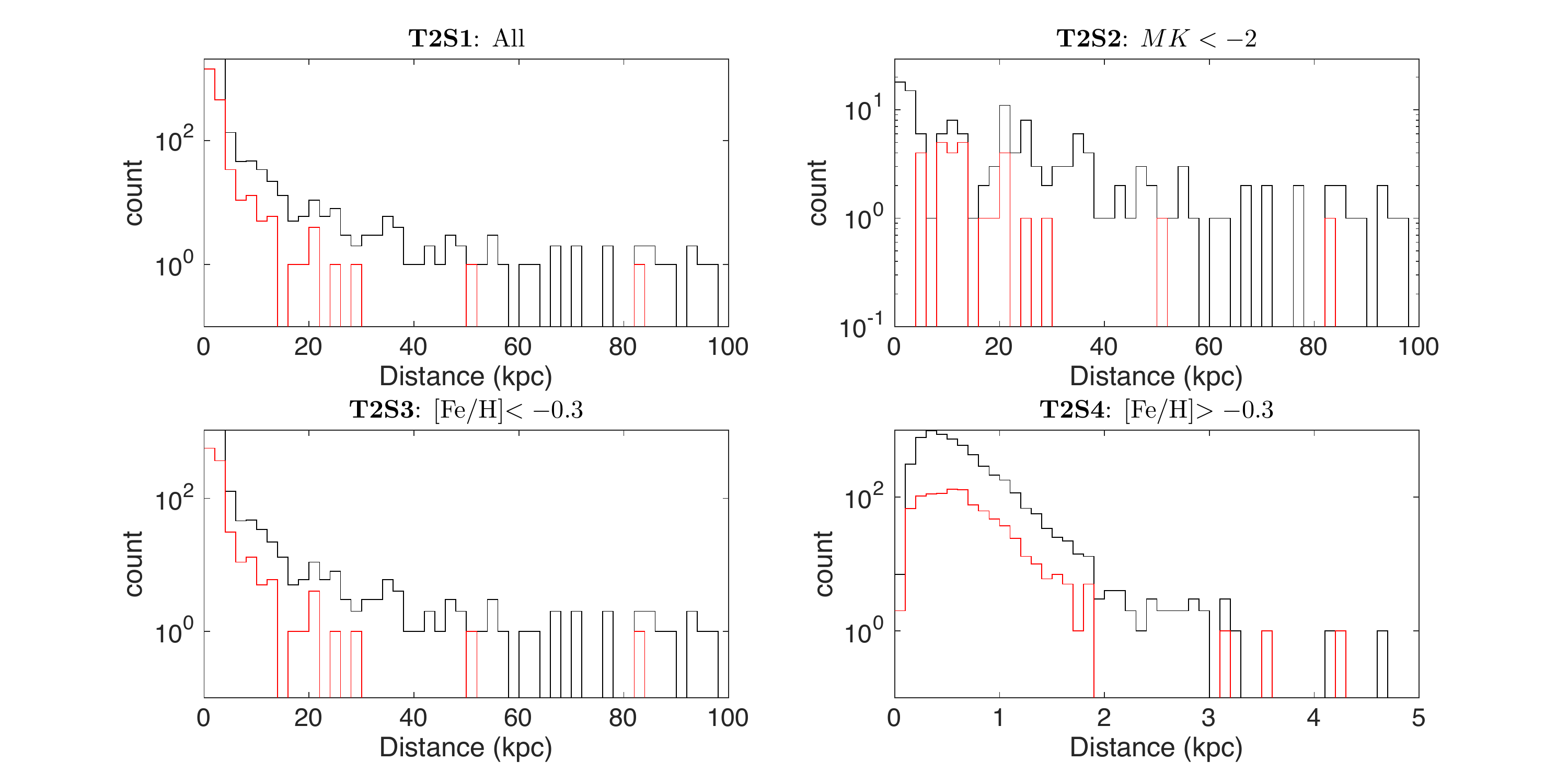}
\caption{Similar to Fig.~\ref{fig:distdistribution1}, but with the selection function \textbf{T2}.}\label{fig:distdistribution2}
\end{figure*}

\begin{figure*}[htbp]
\centering
\begin{minipage}{18cm}
\centering
\includegraphics[scale=0.4]{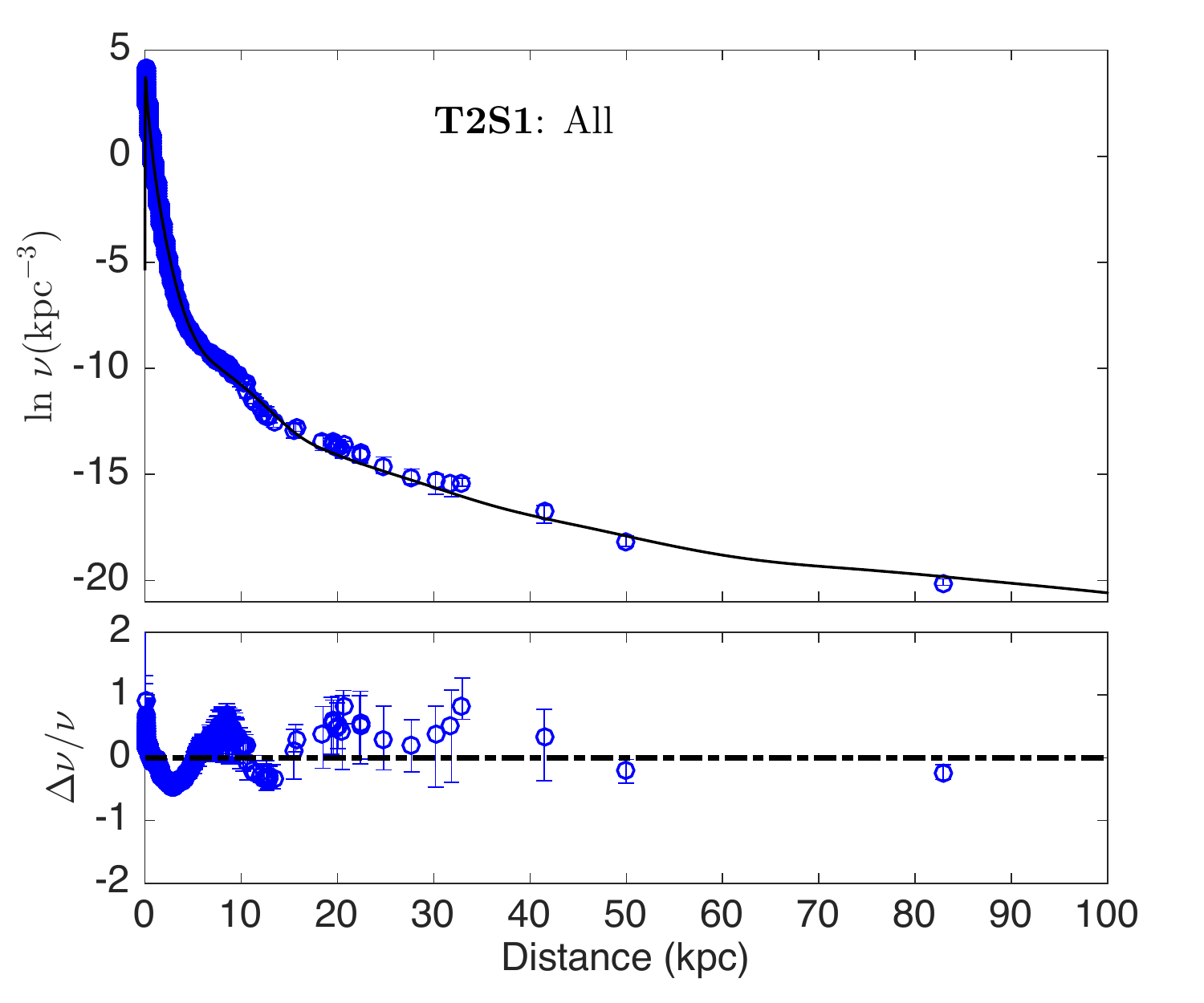}
\includegraphics[scale=0.4]{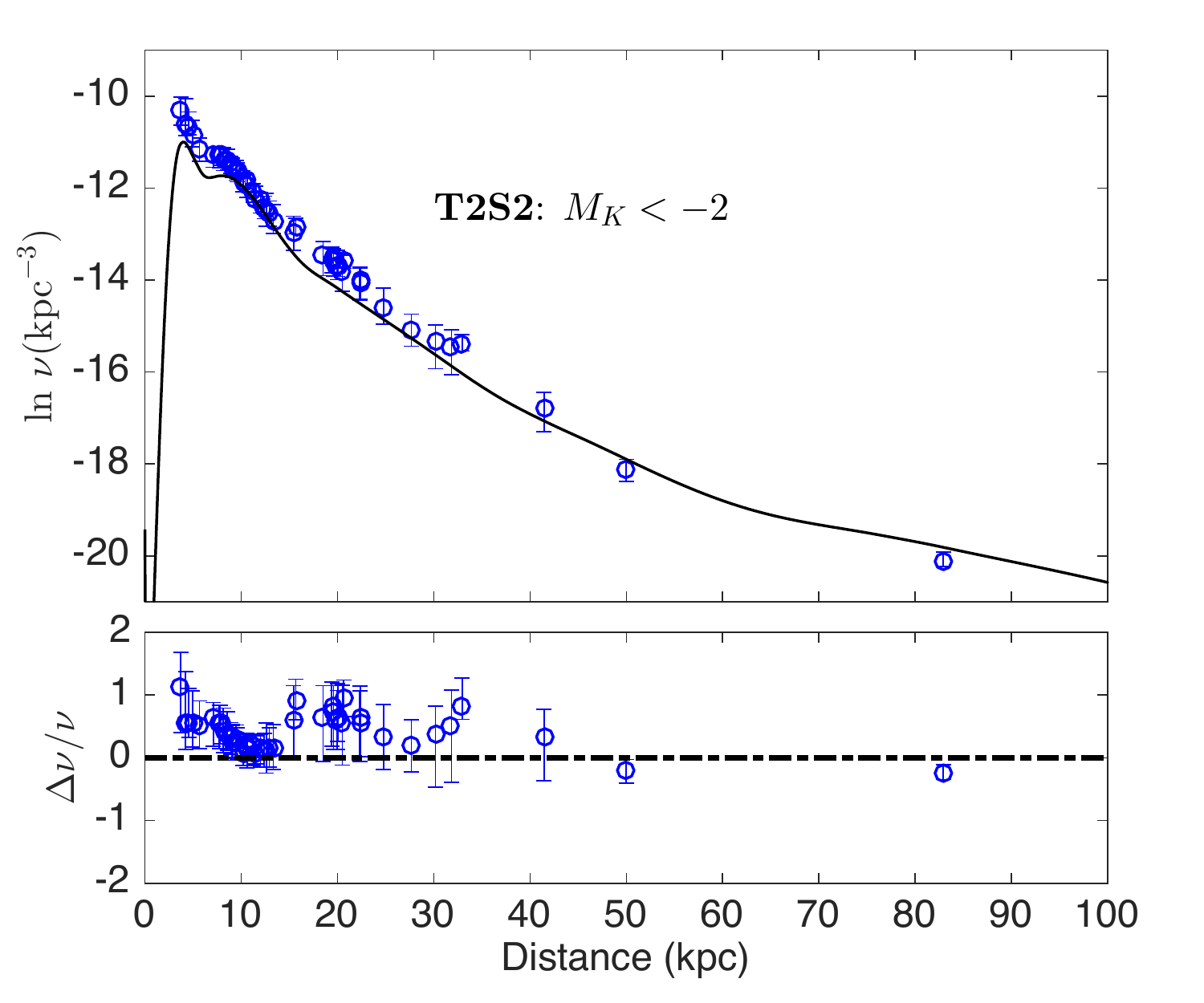}
\includegraphics[scale=0.4]{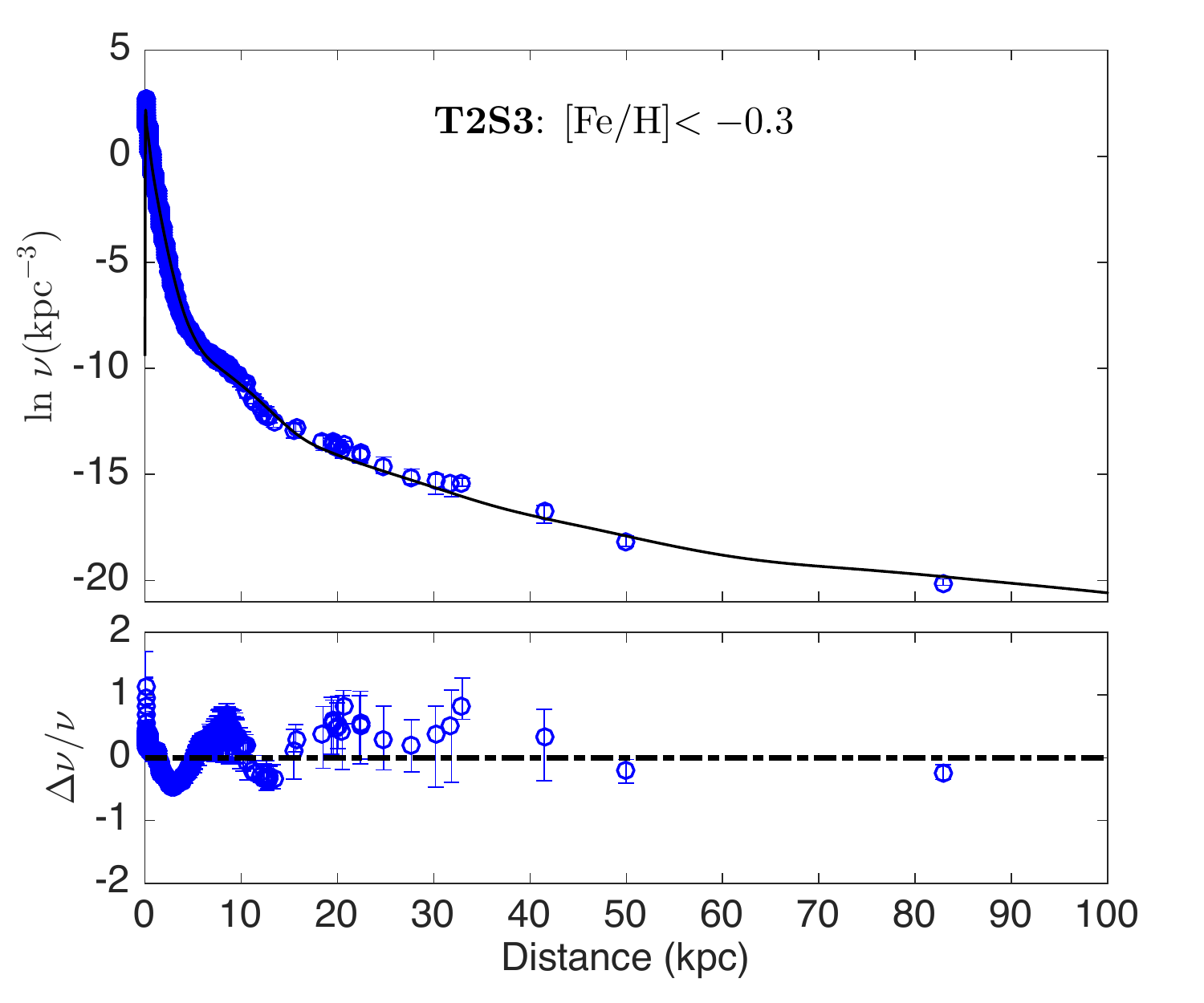}
\includegraphics[scale=0.4]{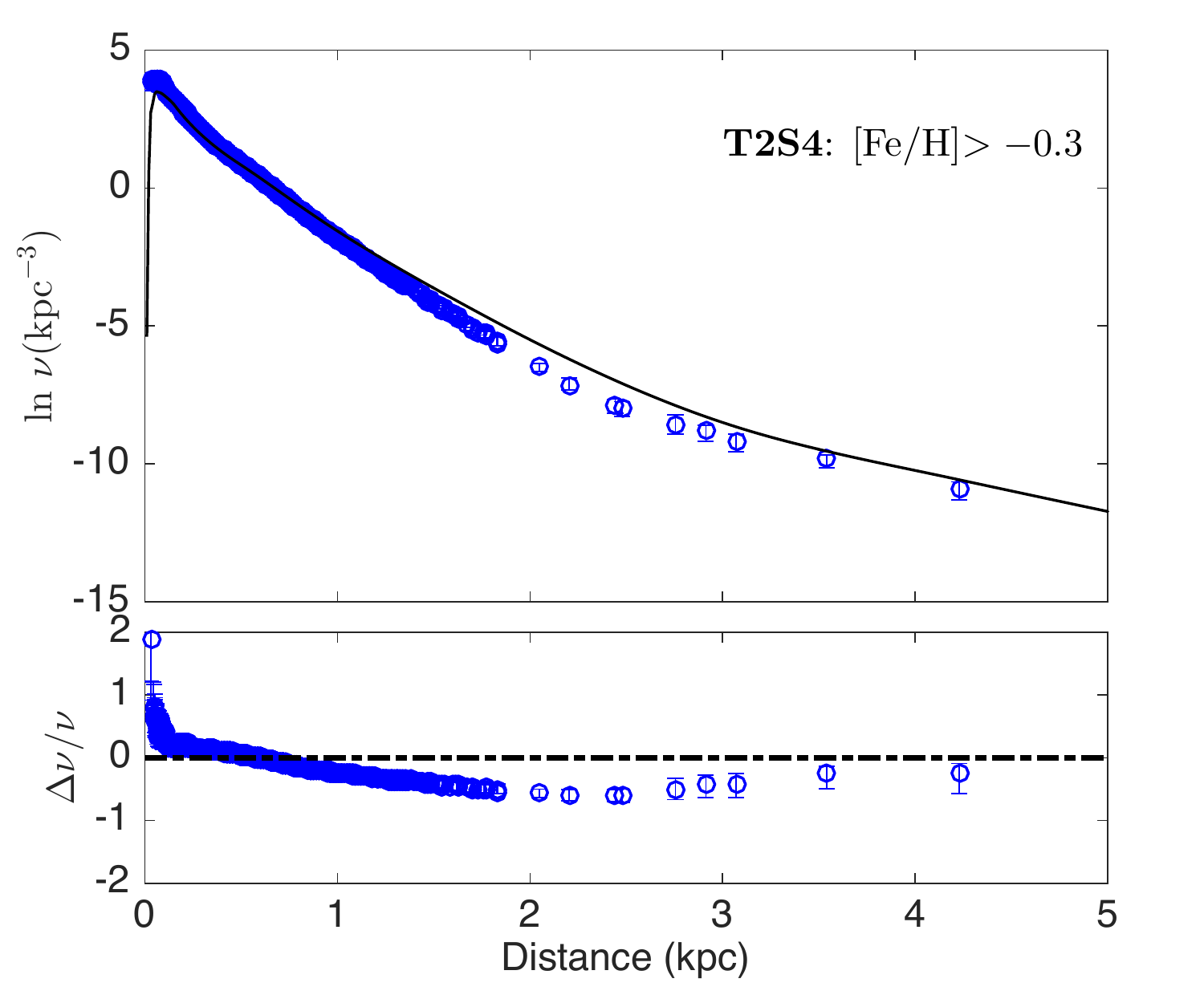}
\end{minipage}
\caption{Similar to Fig.~\ref{fig:nu1}, but with the selection function \textbf{T2}.}\label{fig:nu2}
\end{figure*}

We then change to a more complicated selection function. Because the LAMOST survey selects its targets using optical-band photometry, the bright/faint limiting magnitude cuts in an optical band lead to a sloped cut in $K$ magnitude. Thus a rectangle region selected in optical color-magnitude diagram turns out to be a wedge region in IR-bands. Fig.~\ref{fig:LMS} shows the selection function of a sample plate\footnote{A plate in the LAMOST survey refers to an observed pointing along a line-of-sight. LAMOST has a field-of-view of 20 square degrees. Thus a plate covers the same solid angle. Because LAMOST contains 4000 fibers, a plate can simultaneously observe 4000 objects at most. In practice, a typical number of targets in a plate is around 2500, with some fibers dedicated to observe the sky and flux standard stars, plus a few fibers that are broken and a small fraction of fibers that are not used.} of the LAMOST survey. The middle panel shows that the spectroscopic stars are distributed in a wedge region in the $J-K$ vs. $K$ diagram. Thus the selection function $S^{-1}$ in the right panel also shows the similar shape.

To mimic such a selection function, we randomly select the stars satisfying the following criterion:
%gal1.K>((9-12)/(1-0)*(gal1.JK-0)+12) &...
%    gal1.K<((12-15)/(1-0)*(gal1.JK-0)+15)
\begin{equation}\label{eq:T2}
	-3(J-K)+12<K<-3(J-K)+15.
\end{equation}
We denote this selection function as \textbf{T2}. Fig.~\ref{fig:cmd2} shows a randomly drawn sample based on \textbf{T2} in the $J-K$ vs. $K$ diagram, in which the selected spectroscopic stars (red circles) are distributed in a wedge area. The median numbers of the stars and their scatters over the 50 groups of arbitrarily drawn mock spectroscopic stars based on \textbf{T2} are separately listed in the last column of Table~\ref{tab:sample} for the 4 sub-populations. Fig.~\ref{fig:cmdnumbers2} shows the density of stars in the color-magnitude diagram for the same sample group of mock spectroscopic data as in Fig.~\ref{fig:cmd2}. It is seen that the selected stars show a tilted distribution in the $J-K$ vs. $K$ plane, and so does the map of $S^{-1}$ according to the definition of \textbf{T2}.

Fig.~\ref{fig:distdistribution2} shows the spatial distributions of the star counts without selection function correction for the 4 sub-populations from one of the 50 groups of mock spectroscopic stars (red lines). In contrast, the black lines indicate the distributions of the stars for the corresponding complete datasets. It can be seen that for \textbf{T2S1}, \textbf{T2S2}, and \textbf{T2S3}, distances further than 30\,kpc are significantly under-sampled.  

Fig.~\ref{fig:nu2} shows the derived median stellar density profiles and their 1$\sigma$ dispersions over the 50 randomly drawn groups (blue circles with error bars). In general, the derived profiles are quite consistent with their corresponding ``true'' values within uncertainty of $\Delta\nu/\nu<1$. However, a few systematic biases of about $\Delta\nu/\nu\sim0.5$--1 appear in the derived density profile: one occurs at the distance of $\sim20-30$\,kpc for \textbf{T2S1}, \textbf{T2S2}, and \textbf{T2S3}, the other is at $D\sim2.5$\,kpc for \textbf{T2S4}.

Considering that properties of the Galactic spatial structures, such as the scale height/length of the disks and the power-law index of the stellar halo, are mostly measured in logarithmic density, such systematic biases may not bring severe effects. For instance, a simple exponential fit within 0.3--2.5\,kpc gives a scale height of 219\,pc with the true density profile of \textbf{T2S4} (black line in the bottom-right panel of Fig.~\ref{fig:nu2}), while it decreases to 191\,pc obtained from the corresponding derived stellar density profile (blue circles in the same panel). This means that the derived density based on \textbf{T2} may induce a systematically smaller exponential scale height by only $\sim13$\%. 

For a realistic survey such as LAMOST, many plates are overlapped with each other. And the selection function for different plates varies. Then the possible systematic bias of the stellar density measurement in one plate, as shown in Fig.~\ref{fig:nu2}, may be compensated by other overlapped plates. Consequently, the cummulation of the systematic bias from each individual plate may not induce the overall systematic bias to the final density profiles, but may increase the random uncertainty in the resulting density profiles. 

\section{The stellar density profile of the Milky Way}
\label{sect:density}
\subsection{The selection function of the LAMOST survey}\label{sect:selection}
In general, the targeting strategy of the LAMOST survey is rather simple. Carlin et al. (\cite{carlin2012}) has designed an elegant targeting algorithm, which tries to make the distribution of the target stars flat in color-magnitude planes. However, in practice, this technique has not been fully used for a few reasons. First, the dynamical range of magnitudes observed in a single LAMOST plate is limited to only about 3 magnitudes so that bright spectra will not saturate or cross-talk between fibers. Second, the actual limiting magnitude of $r\sim18$\,mag is brighter than the designed goal by at least 2\,mag. Both situations lead to insufficient source stars for targeting. For instance, the bright plates for LAMOST survey cover $14<r<16.8$\,mag, which may only contain less than 200 stars per square degree at high Galactic latitudes. Such a sparse stellar sampling is not sufficient to apply the selection function designed by Carlin et al. (\cite{carlin2012}) for targeting, because the targeting algorithm needs far more source stars than available fibers to achieve a flat distribution of targets in color-magnitude plane. 

Finally, the LAMOST survey separated the targets into different plates with different ranges of magnitudes at each line-of-sight. The \emph{VB} plates cover $9<r<14$, the \emph{B} plates cover $14<r<16.8$, the \emph{M} plates cover $r<17.8$, and the \emph{F} plates cover $r<18.5$. For the \emph{VB} and \emph{B} plates, no specific selection function was applied, i.e., the stars are randomly selected. For the \emph{M} and \emph{F} plates, we only applied the selection function in $r$ magnitude, and leave the selection in color index to be arbitrary. 

This simplified selection function was applied to all of the sky area observed by LAMOST except for the Galactic anti-center region, which covers $150^\circ<l<210^\circ$ and $|b|<30^\circ$. For the main survey regions, the UCAC4 catalog (Zacharias et al. \cite{zacharias2013}) with $r<14$\,mag and PanSTARRS-1 catalog (Schlafly et al. \cite{schlafly2012}) with $r>14$ were adopted as the source catalogs for targeting. For the Galactic anti-center region, the selection strategy follows Yuan et al. (\cite{yuan2015}), and uses 2MASS, UCAC4, and Xuyi survey catalogs (Liu X.-W. et al. \cite{liux2014}) as the sources for targeting.

Because the different source catalogs are based on different photometric systems, they can not be unified unless they are cross-calibrated with each other. To avoid the complicated calibrations between at least 4 different systems, we finally adopt the 2MASS catalog, which covers most of the LAMOST observed stars, as the photometric dataset. Indeed, for the LAMOST K giant stars, about 96\% of them are within $K<14.3$\,mag, which is the limiting magnitude of the 2MASS catalog. %For the 2MASS catalog, the stars are roughly complete to its limiting magnitude, which is $K=14.3$.  Most of the LAMOST stellar targets are brighter than the limiting magnitude of the 2MASS photometry. 
Subsequently, we use $J-K$ vs. $K$ map to derive the selection function $S$ for each observed plate of the LAMOST survey. The selection function $S^{-1}$ for all LAMOST DR3 plates and the \emph{python} codes used for the measurement of the stellar density profiles can be found at \href{https://github.com/liuchaonaoc/LAMOST_density}{https://github.com/liuchaonaoc/LAMOST\_density}.

\subsection{LAMOST RGB stars}\label{sect:RGB}
In order to derive the density profiles for the stellar disk and halo of the Galaxy, we select red giant branch (RGB) stars. Although RGB stars are not the standard candles as the  red clump stars, the accuracy of the distance for the former ones  is still around 20-30\%~(Liu et al. \cite{liu2014}, Carlin et al. \cite{carlin2015}) is sufficient for mapping Galactic structure in the outer regions. Moreover, RGB stars are intrinsically brighter and occur in stellar populations with a broader range of metallicity than red clump stars. Therefore, they are suitable tracers for the outer disk and the halo, which are dominated by the metal-poor stars.

For LAMOST survey data, we adopt the stellar parameters, i.e., the effective temperature, surface gravity, and metallicity, provided by the LAMOST pipeline (Wu et al. \cite{wu2011}, \cite{wu2014}, Luo et al.~\cite{luo2015}). The K giant stars, including the RGB and red clump stars, are selected based on the criteria proposed by Liu et al. (\cite{liu2014}). Red clump stars are identified by Wan et al. (\cite{wan2015}) and Tian et al. (\cite{tian2016}). The RGB stars are then selected from the K giant stars by excluding the red clump stars. 

\begin{figure*}[htbp]
\includegraphics[scale=0.6]{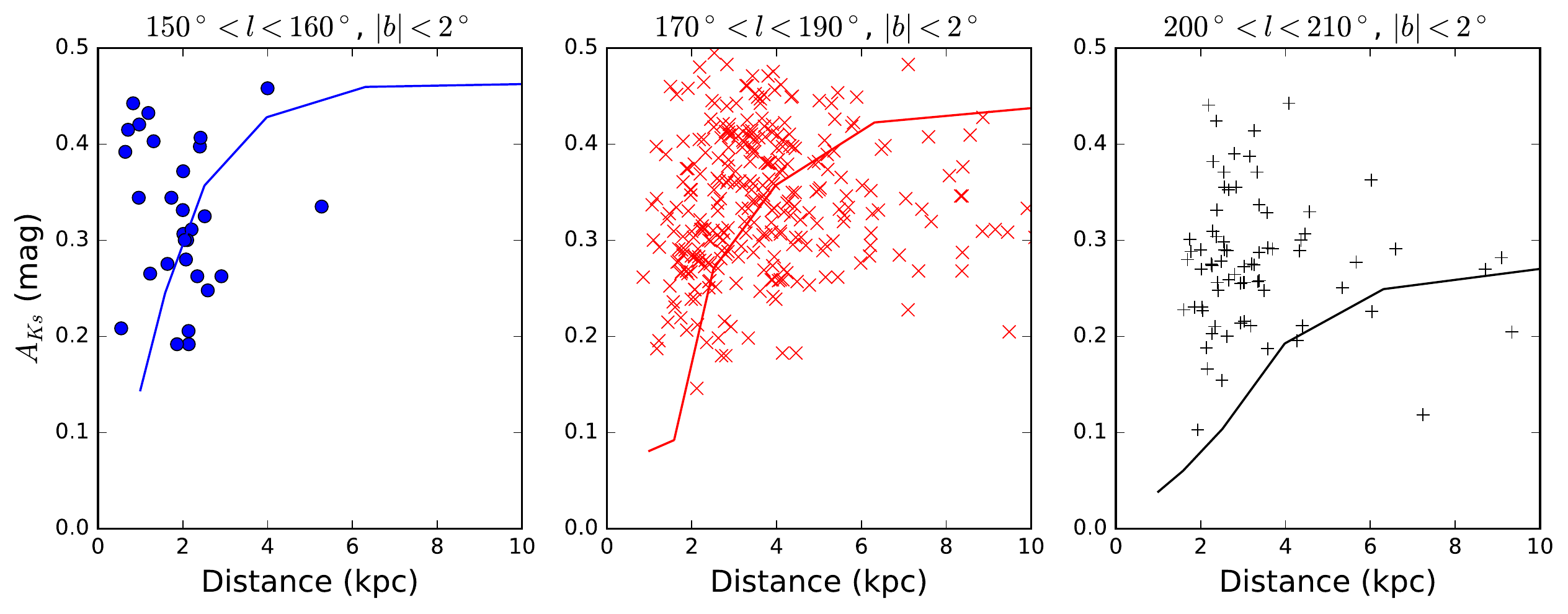}
\caption{The blue circles, red crosses, and black plus symbols indicate the RGB stars with $|b|<2^\circ$ along the lines-of-sight $150^\circ<l<160^\circ$, $170^\circ<l<190^\circ$, and $200^\circ<l<210^\circ$, respectively. Three extinction-distance relations at $l=150^\circ$ (blue line), $l=180^\circ$ (red line), and $l=210^\circ$ (black line) queried from the 3-D extinction map of Green et al. (\cite{green2014}, \cite{green2015}) are superposed on the observed data.}\label{fig:Ak}
\end{figure*}

The absolute magnitudes at $K_s$ band, $M_{Ks}$, for the RGB stars are estimated following the method suggested by Carlin et al. (\cite{carlin2015}). Interstellar extinction for the individual RGB stars is estimated using the Rayleigh-Jeans color excess (RJCE) approach introduced by Majewski et al. (\cite{majewski2011}) and later revised to incorporate WISE bandpasses by Zasowski et al. (\cite{zasowski2013}). Subsequently, the distances to the RGB stars are estimated by combining the apparent $K_s$ magnitude, the absolute magnitude $M_{Ks}$, and the RJCE extinction $A_{Ks}$. 

The RJCE extinction is compared with the 3-D extinction map provided by Green et al. (\cite{green2014}, \cite{green2015}), who derived the 3-D dust map from the PanSTARRS-1 data. Fig.~\ref{fig:Ak} shows the RJCE extinction vs. distance for the RGB stars located at ($150^\circ<l<160^\circ$,  $|b|<2^\circ$), ($170^\circ<l<190^\circ$,  $|b|<2^\circ$), and ($200^\circ<l<210^\circ$,  $|b|<2^\circ$) using blue circles, red crosses, and black pluses, respectively. The Green et al. mean extinction-distance relations at the  similar lines-of-sight are superposed in the same figure. We see that the RJCE extinctions for individual RGB stars are roughly consistent with Green's 3-D extinction map within about 0.2\,mag, which may lead to only about 10\% uncertainty in distance. Therefore, the distances for the RGB stars should not be substantially affected by extinction, even in the Galactic mid-plane.

In the following subsections we demonstrate two samples: 1) RGB stars with all ranges of metallicity as the tracers for the Milky Way structures, especially for the  Galactic disk, and 2) metal-poor RGB stars as tracers for the stellar halo.

\begin{table}
\caption{Explanation of data columns in the catalogs used for disk and halo density mapping.}\label{tab:datamodel}
\begin{tabular}{l|c|c|l}
\hline\hline
column name & type & unit & comments\\	
\hline
obsid & integer & & The identifier of the LAMOST spectra\\
RA & double & degree & Right ascension in epoch J2000.0\\
DEC & double & degree & Declination in epoch J2000.0\\
%$l$ & double & degree & Galactic longitude\\
%$b$ & double & degree & Galactic latitude\\
$T_{\rm eff}$ & float & K & Effective temperature from LAMOST DR3 catalog\\
$\log g$ & float & dex & Surface gravity from LAMOST DR3 catalog\\
$\rm [Fe/H]$ & float & dex & Metallicity from LAMOST DR3 catalog\\
$v_{los}$ & float & km\ s$^{-1}$ & Line-of-sight velocity from LAMOST DR3 catalog\\
%$EW_{Mgb}$ & float & $\rm\AA$ & Equivalent width of Mg$_b$ line\\
$M_{Ks}$ & float & mag & $K_s$-band absolute magnitude from Carlin et al. (\cite{carlin2015})\\
$M_{Ks}$ lower error & float & mag & $-1\sigma$ error of $M_{Ks}$\\
$M_{Ks}$ upper error & float & mag & $+1\sigma$ error of $M_{Ks}$\\
$A_{Ks}$ & float & mag & $K_s$-band interstellar extinction derived from RJCE (Majewski et al. \cite{majewski2011},\\
& & & Zasowski et al. \cite{zasowski2013})\\
$K_s$ & float & mag & $K_s$-band apparent magnitude from 2MASS catalog\\
Distance & float & kpc & Distance estimated from $M_{Ks}$, $A_{Ks}$, and $K_s$\\
Distance lower error & float & kpc & $-1\sigma$ error of distance\\
Distance upper error & float & kpc & $+1\sigma$ error of distance\\
$Z$ & float & kpc & Vertical distance from the star to the Galactic mid-plane\\
$R$ & float & kpc & Galactocentric radius in cylindrical coordinates\\
$r$ & float & kpc & Galactocentric radius in spherical coordinates\\
$\ln\nu$ & float & $\ln$\ kpc$^{-3}$ & Logarithmic stellar density at the position of the star\\
\hline\hline
\end{tabular}	
\end{table}

\begin{table*}\tiny
\centering
\caption{The disk-like RGB stars selected from the LAMOST DR3 catalog. (Here we only list the first 3 rows, the complete table is in the on-line file).}\label{tab:diskRGB}
\begin{tabular*}{1.05\textwidth}[-2cm]{@{}c m{0.7cm} m{0.7cm} m{0.5cm} m{0.5cm} m{0.5cm} m{0.35cm} cccccccl}%{1.2\textwidth}{XXXXXXXXXXXXXXX}%
	\hline\hline
	obsid & RA & DEC & $T_{\rm eff}$ & $\log g$ & [Fe/H] & $v_{los}$ & $M_K$ & $A_K$ & $K$ & Distance & $Z$ & $R$ & $r$ & $\ln\nu$\\
	& deg & deg & K & dex & dex & km\ s$^{-1}$ & mag & mag & mag & kpc & kpc & kpc & kpc & $\ln$(pc$^{-3}$)\\ 
	\hline
75606117 & 337.46744 & +6.01526 & 4829 & 1.11 & -1.66 & -3 & -3.96$_{-0.44}^{+0.42}$ & 0.148 & 10.646 & 7.79$_{-1.37}^{+1.75}$ & -5.22 & 8.26 & 9.77 & -11.40\\
269507178 & 3.66002 & +55.72871 & 4159 & 1.47 & -0.47 & -93 & -3.82$_{-0.36}^{+0.24}$ & 0.252 & 9.721 & 4.55$_{-0.48}^{+0.82}$ & -0.51 & 10.86 & 10.88 & -6.25\\
51211049 & 257.70419 & +18.38338 & 4207 & 2.64 & -0.51 & -33 & -3.32$_{-0.42}^{+0.38}$ & 0.072 & 9.687 & 3.86$_{-0.62}^{+0.82}$ & 1.97 & 5.82 & 6.14 & -7.51\\
\hline\hline
\end{tabular*}
\end{table*}

\subsection{The stellar disk}\label{sect:disk}
\begin{figure}[htbp]
\begin{minipage}{9cm}
\includegraphics[scale=0.7]{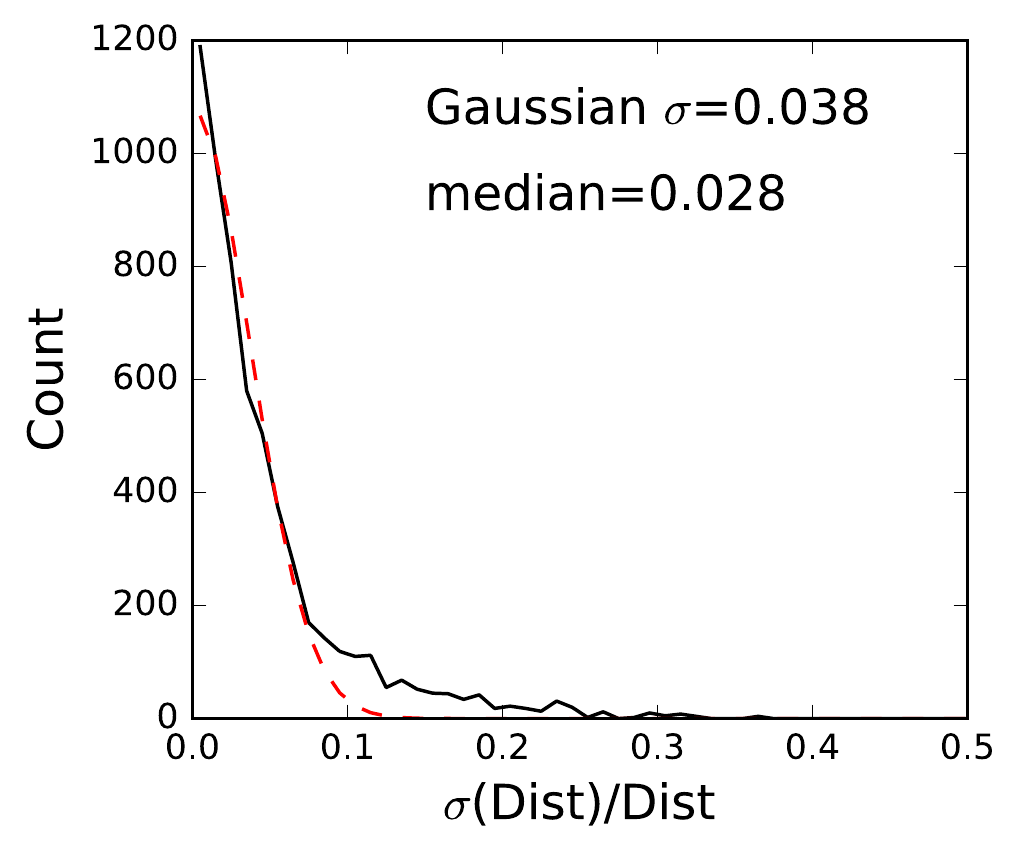}	
\includegraphics[scale=0.7]{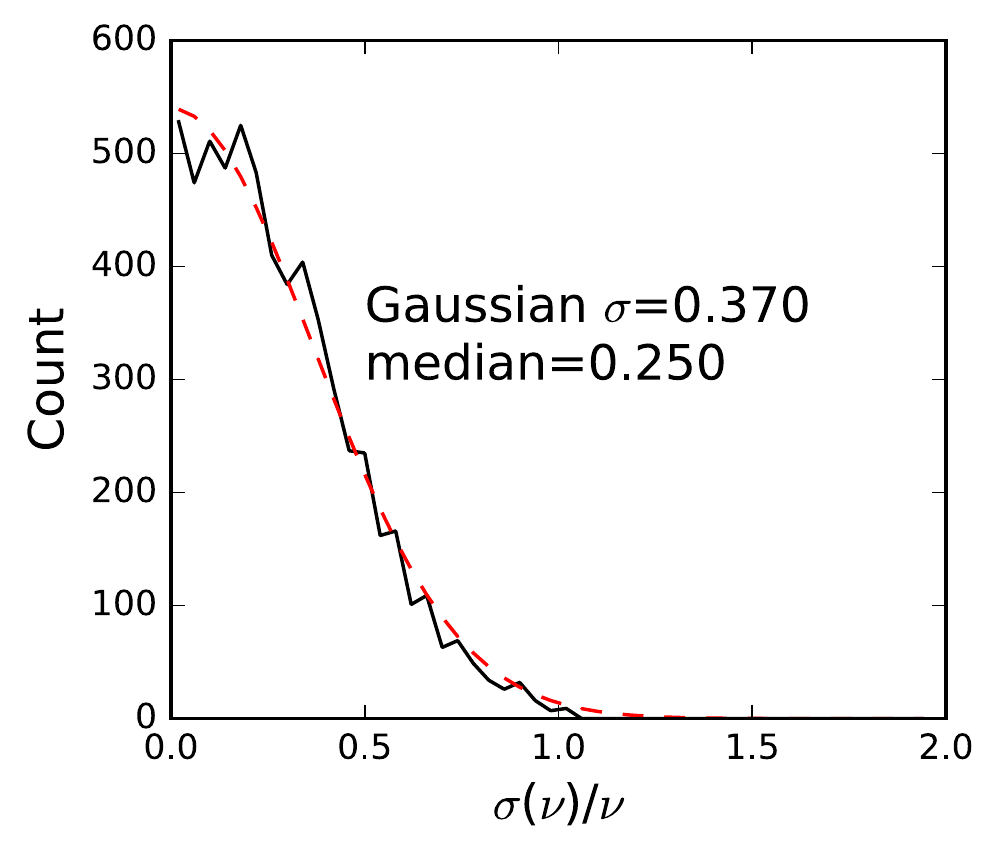}
\end{minipage}
\caption{In the top panel, the black line shows the distribution of the relative scatter of the distance for the repeatedly observed RGB stars. The red dashed line is the best fit Gaussian of the distribution of the scatter. The $\sigma$ of the Gaussian is $0.038$ and the median of the scatter is $0.028$. In the bottom panel, the black line displays the relative scatter of the $\nu$ for the repeatedly observed RGB stars. The red dashed line is the best fit Gaussian with $\sigma=0.37$. The median of the scatter of $\nu$ is $0.25$.}\label{fig:deltanu}
\end{figure}

\begin{figure}[htbp]
\includegraphics[scale=0.7]{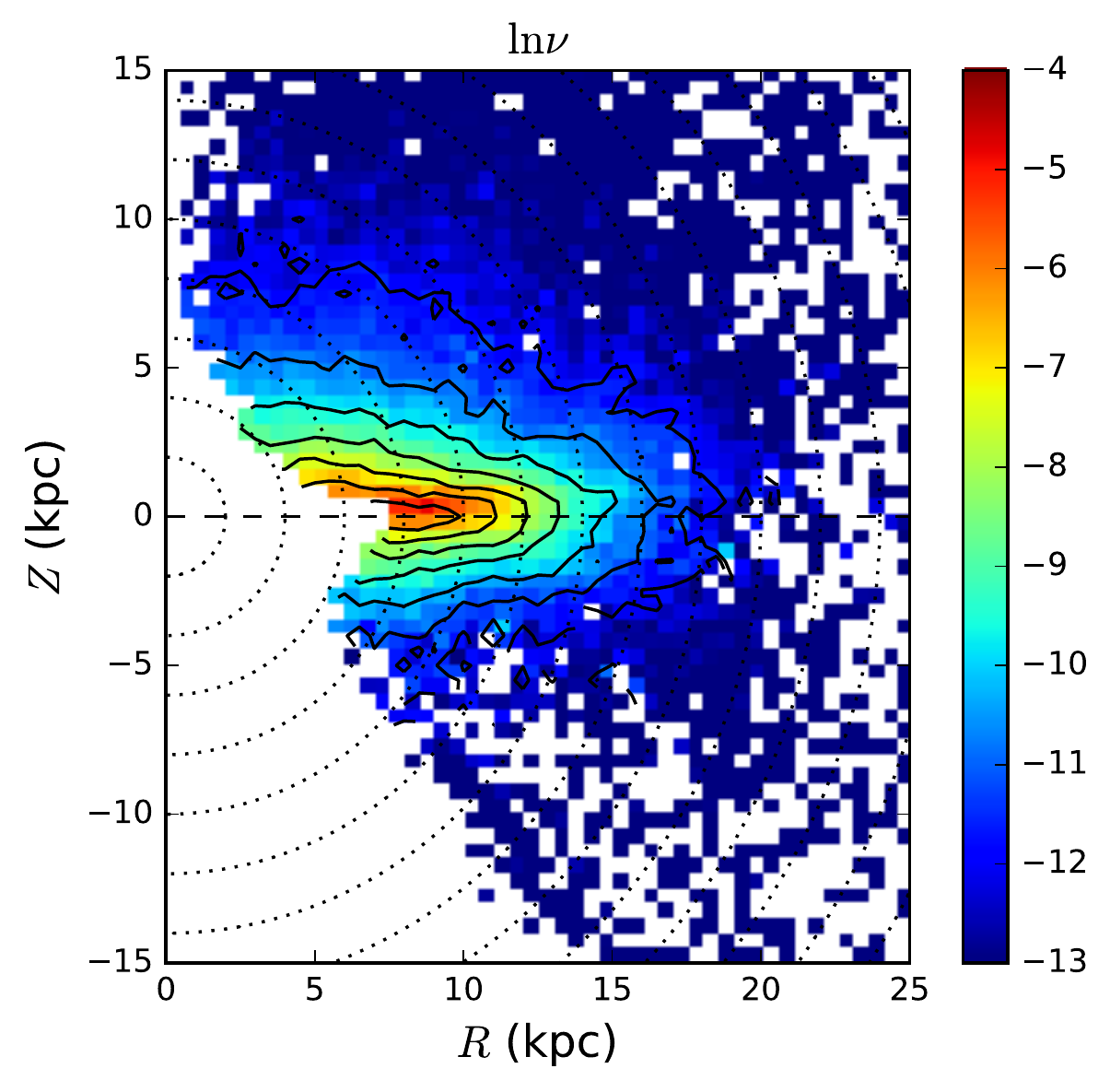}
\caption{The figure shows the averaged stellar density map for the RGB stars with all range of metallicity, $M_{Ks}<-3.5$, and $D>0.5$\,kpc. The color codes the mean $\ln\nu$ for each $R$--$Z$ bin. The black contours indicate the iso-surface lines at $\ln\nu=-12$, $-11$, $-10$, $-9$, $-8$, and $-7$ from outside to inside, respectively. The dotted circles indicate the Galactocentric radii of 2, 4, ..., 24\,kpc from left to right, respectively. The black horizontal dashed line indicates the plane of $Z=0$.}\label{fig:nuDisk}
\end{figure}

\begin{figure}[htbp]
\includegraphics[scale=0.7]{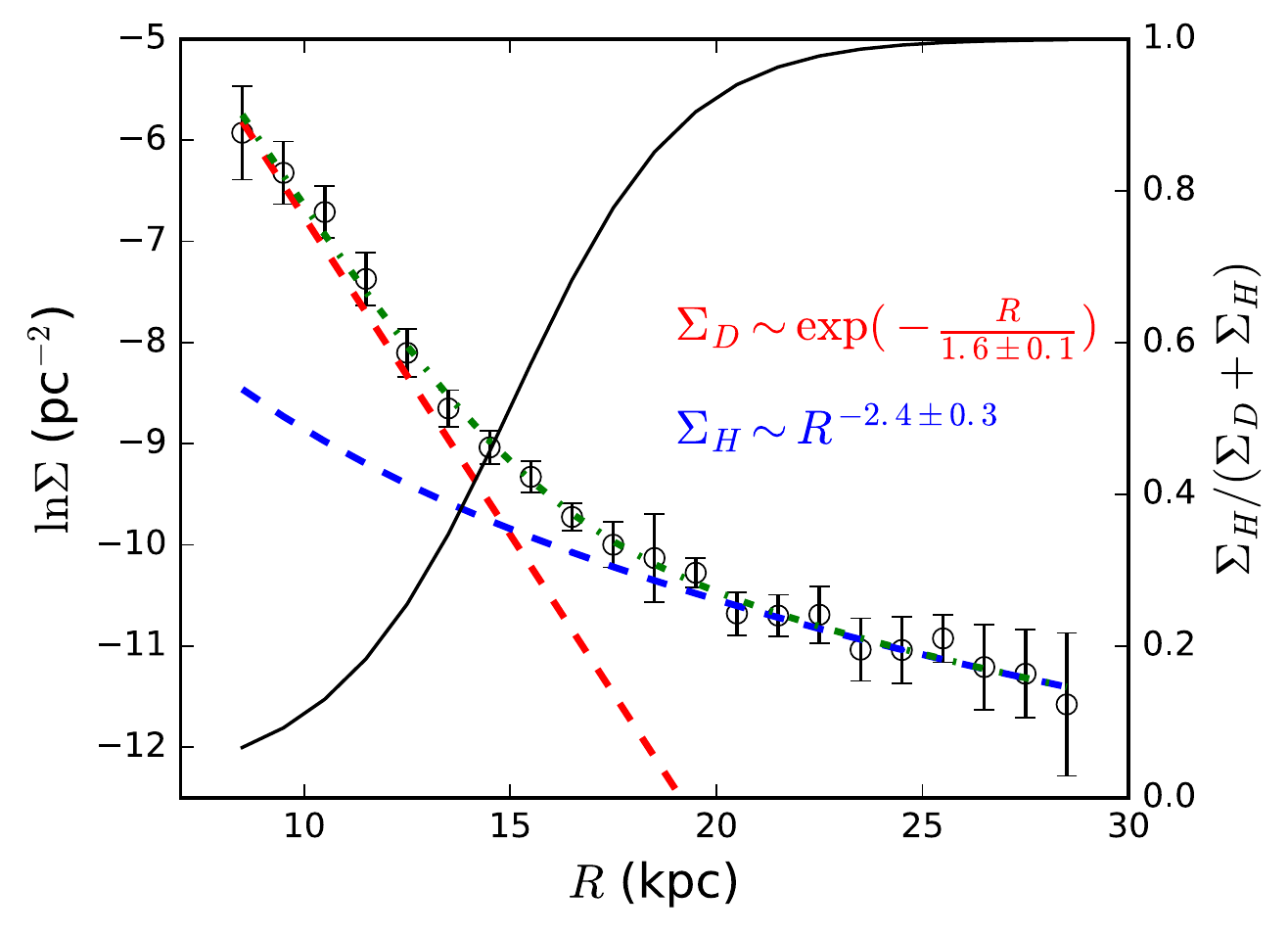}
\caption{The black circles display the radial stellar surface density profile integrated over $Z$ for the RGB stars. The bin size of $R$ is 1\,kpc. The red and blue dashed lines stand for the best fit exponential disk and power-law halo profiles, respectively. The scale length for the disk is $1.6\pm0.1$\,kpc and the power-law index for the halo is $2.4\pm0.3$. The green dashed line is the sum of the two model profiles for the halo and disk. The black line aligned with the right $y$-axis shows the fraction of the halo relative to the total surface density as a function of $R$.}\label{fig:diskprofile}
\end{figure}

We select 21,954 RGB stars with $M_{Ks}<-3.5$\,mag and distances larger than 0.5\,kpc as probes to map the Galaxy. Because these samples are dominated by the metal-rich stars, they mostly trace the Galactic disk. The volume completeness for these RGB stars are within 40\,kpc. The volume completeness in this paper is defined such that the range of the luminosity of the stars does not significantly change within the volume. Plate by plate, we apply the method to derive the stellar density and finally obtain density values at the  distances at which sample stars are located. Table~\ref{tab:diskRGB} lists the first three rows of the samples. The columns in the table are described in Table~\ref{tab:datamodel}. The full RGB data can be found at \href{https://github.com/liuchaonaoc/LAMOST_density}{https://github.com/liuchaonaoc/LAMOST\_density}.

Note that the samples contain about 20\% of stars that have been observed more than twice. Because the duplicated stars were observed in different plates and the selection effects for each observation can be considered separately, the duplication would not affect the stellar density estimation. In fact, these data can be used to test the internal uncertainty, or the precision, of the distance and stellar density estimations, respectively. The top panel of Fig.~\ref{fig:deltanu} shows the distribution of the relative scatter of the distance estimates (black line) for the stars observed multiple times. The scatter is defined by the standard deviation of the distance. The relative scatter is the scatter divided by the mean distance estimate. The distribution of the relative scatter of distance is fitted with a Gaussian (red dashed line) with the best fit parameter of $\sigma=0.038$, while the median value of the distribution is $0.028$. This means that the internal uncertainty of the distance estimates for the RGB stars is around $3-4$\%.

The bottom panel of Fig.~\ref{fig:deltanu} shows the distribution of the relative scatter of stellar density (black line) for the duplicated stars. The definitions of the scatter and relative scatter are similar to those for the distance estimates. The red dashed line stands for the best fit Gaussian with $\sigma=0.37$. The median value for the distribution is $0.25$. This is better than the performance of the mock data test shown in Sect.~\ref{sect:T1} and~\ref{sect:T2}, which may reach $\Delta\nu/\nu>0.5$ in the worst cases. The difference in the derived stellar densities for the multiple-observed stars is originated from two possible channels: 1) the different selection corrections from the different plates observing the same star and 2) the different averaging sky areas since the plates observing the same star may not be completely overlapped with each other. The small dispersion of the relative scatter implies that these differences are very small for most of the stars.
 
Fig.~\ref{fig:nuDisk} demonstrates the averaged stellar density map in $R$--$Z$ plane\footnote{$R$ and $Z$ are Galactocentric cylindrical coordinates with adopted solar position at $R_0=8$\,kpc and $Z_0=0.027$\,kpc (Chen et al. \cite{chen2001})}. The bin size is 0.5$\times$0.5\,kpc. Because this population is dominated by metal-rich stars, it displays a remarkable tomographic map of the Galactic disk from $R\sim4$ to $\sim20$\,kpc. Two prominent features are seen in this figure. First, the Monoceros ring, unveiled by Newberg et al. (\cite{newberg2002}) and later highlighted in SDSS star counting (Juri\'c et al. \cite{juric2008}) and metallicity distribution mapping (Ivezi{\'c} et al.\cite{ivezic2008}), does not show up in our map of the whole Galactic outer disk with LAMOST RGB stars. Second, the iso-density contours (black lines) show  moderate north-south asymmetry in the outer part of the disk starting from $R\sim12$\,kpc. The stellar density above the mid-plane is larger than that below at given radii. This is essentially consistent with Xu et al. (\cite{xu2015}). The asymmetry about the mid-plane may also be the result of the warp. If that is the case, the stellar disk may bend up to some extent in the Galactic anti-center region between $150^\circ<l<210^\circ$, in which the LAMOST disk data are mostly concentrated. Both features are very interesting and will be further discussed in an upcoming paper (Wang et al. in preparation).

%Discuss the Monoceros ring here.

%Discuss the asymmetry here.

In Fig.~\ref{fig:diskprofile}, the black circles show the stellar surface density derived by integrating the volume density over $Z$ in each $R$ slice. In order to avoid issues due to the lack of data in the southern Galactic hemisphere, we use $|Z|$ instead of $Z$. That is, the mean stellar density at given $|Z|$ is averaged over the values at both $+Z$ and $-Z$. Only  $|Z|<40$\,kpc are included in the integration. The width of each $R$ slice is 1\,kpc. For some bins without stellar density estimates, interpolated values are used. The red and blue dashed lines are best-fit exponential disk and power-law halo profiles, respectively, and the green dot-dashed line is the sum of the two profiles. We find that the surface density profile can be well fitted with an exponential disk with scale length of $1.6\pm0.1$\,kpc and a power law halo with index of $2.4\pm0.3$  Note that this power-law index is not equivalent to that derived from spherical coordinates, since it is derived from the surface density as a function of the cylindrical radius $R$. 

The fraction of stars contributed by the halo, as derived from our models of the surface density, is indicated as the black line aligned with the right $y$-axis in the figure. It shows that, at $R=14$\,kpc, about half of the surface density is contributed by the disk. Therefore, no evidence is found to support that there is a truncation at around 14\,kpc. Moreover, at $R=19$\,kpc, the disk still contributes $\sim10$\% to the total stellar surface density, which cannot be negligible because the data points (black circles) are clearly larger than the halo model (blue lines) around the radius. In other words, the disk extends to as far as 19\,kpc. Beyond this radius, the observed surface density smoothly transitions to a pure halo profile. The perfect fitting with an exponential disk model implies that the disk displays neither truncation, nor break, nor up-bending feature. 

The extended larger size of the disk is not likely to be an artificial effect due to the incorrect interstellar extinction, according to the discussion in sub-section~\ref{sect:RGB}. On the other hand, Carlin et al. (\cite{carlin2015}) pointed out that the distances may be systematically overestimated for $\alpha$-enhanced giant stars by about 15\%. However, Hayden et al.~(\cite{hayden2015}) shows that it is the $\alpha$-low giant stars who dominate the outer disk. Consequently, the systematic bias in distance due to the $\alpha$ abundance may not significantly affect the distance for the outer disk RGB stars. Therefore, we suggest that the largely extended disk without truncation should be real.

The Galactic disk measured in this work is significantly larger than some other works, which claimed that the Milky Way stellar disk truncates at around 14--15\,kpc (e.g. Robin et al.~\cite{robin1992}, Minniti et al.~\cite{minniti2011}). Carraro~(\cite{carraro2015}) pointed out that the lines-of-sight investigated by these works are very close to $b\sim0$ and may not be aligned with the disk plane in the outskirts due to the existence of the warp. In this case, at some radius the in-plane lines-of-sight no longer probes the disk, leading to a false truncation at that radius. It is worth to note that Carraro et al.~(\cite{carraro2010}) and Feast et al.~(\cite{feast2014}) found some young stars at around 20\,kpc from the Galactic center, which is in agreement with our suggestion that the disk extends to such radii. We also point out that Xu et al.~(\cite{xu2015}) has suggested that the disk may extend all the way to large radii. More quantitative analysis will appear in upcoming work by Wang et al. (in preparation).
\begin{figure}[htbp]
\includegraphics[scale=0.7]{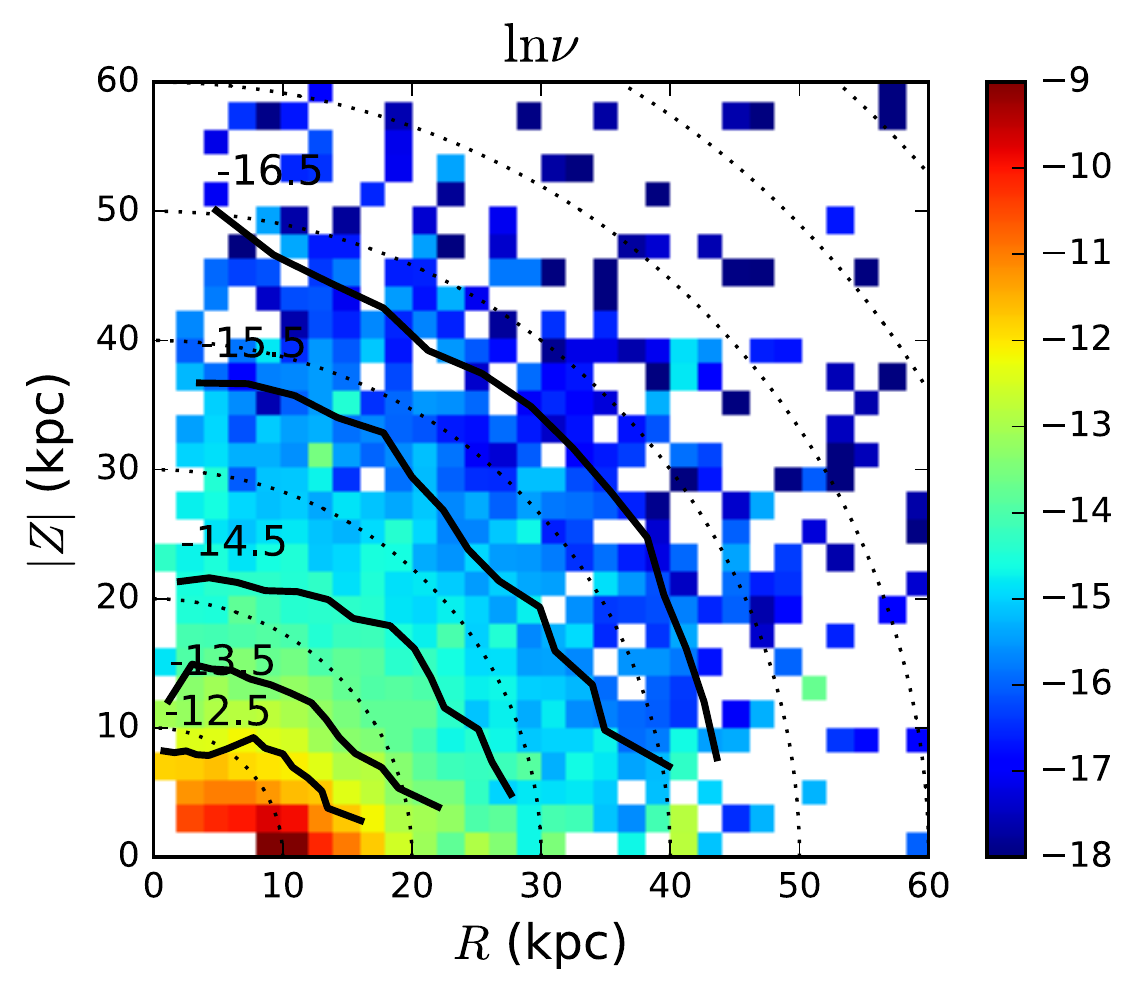}
\caption{The figure shows the averaged stellar density map for the halo RGB stars in the $R$--$Z$ plane. The color codes the mean $\ln\nu$ in each $R$--$Z$ bin. The black contours indicate the iso-surface lines at $\ln\nu=-16.5$, $-15.5$, $-14.5$, $-13.5$, and $-12.5$ from outside to inside, respectively. The dotted circles indicate the Galactocentric radii of 10, 20, ..., 80\,kpc from bottom-left to top-right, respectively.}\label{fig:nuHalo}
\end{figure}

\begin{figure}[htbp]
\includegraphics[scale=0.7]{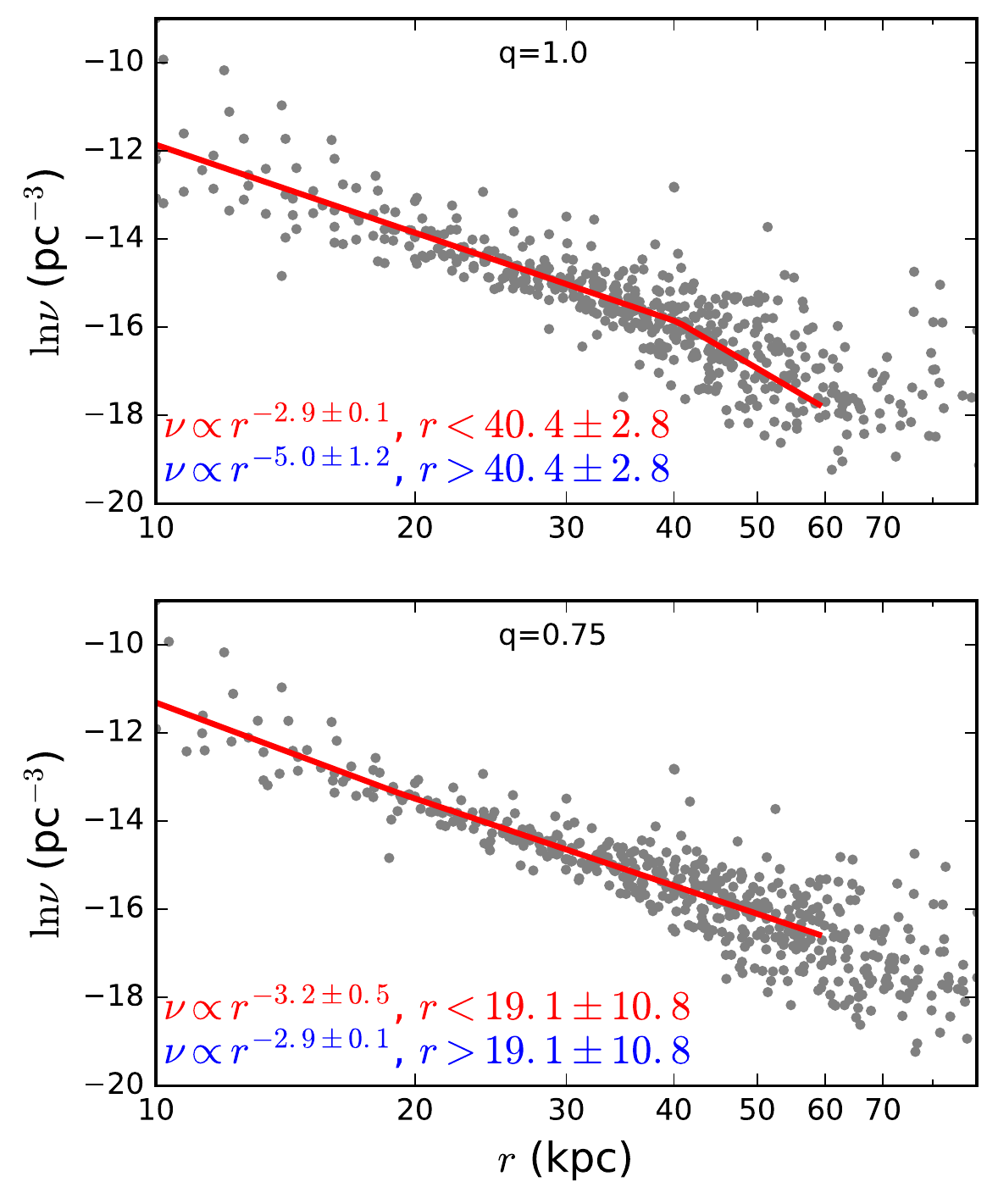}
\caption{In the top panel, with the fixed axis ratio of $q=1.0$, the black dots show the mean logarithmic stellar density values extracted from Fig~\ref{fig:nuHalo} by converting ($R$, $Z$) into radii $r$. The red line is the best fit broken power-law profile with a break radius of $40.4\pm2.8$\,kpc. Within the break radius, the power-law index is $-2.9\pm0.1$, while beyond this radius the index becomes $-5.0\pm1.2$. The bottom panel shows the data with $q=0.75$. In this case the broken power-law does not work quite well in the sense that the break radius has a large uncertainty of $10.0$\,kpc and the inner and outer power-law indices, which are $-3.2\pm0.5$ and $-2.9\pm0.1$ respectively, are quite similar. Therefore, it seems that a single power-law may be more suitable for this case.}\label{fig:haloprofile}
\end{figure}

\subsection{The stellar halo}\label{sect:halo}

\begin{table*}\tiny
\centering
\caption{The halo-like RGB stars selected from the LAMOST DR3 catalog. (Here we only list the first 3 rows, the complete table is in the on-line file).}\label{tab:hRGB}
\begin{tabular*}{1.05\textwidth}[-2cm]{@{}c m{0.7cm} m{0.7cm} m{0.5cm} m{0.5cm} m{0.5cm} m{0.35cm} cccccccl}%{1.2\textwidth}{XXXXXXXXXXXXXXX}%
	\hline\hline
	obsid & RA & DEC & $T_{\rm eff}$ & $\log g$ & [Fe/H] & $v_{los}$ & $M_K$ & $A_K$ & $K$ & Distance & $Z$ & $R$ & $r$ & $\ln\nu$\\
	& deg & deg & K & dex & dex & km\ s$^{-1}$ & mag & mag & mag & kpc & kpc & kpc & kpc & $\ln$(pc$^{-3}$)\\ 
	\hline
75606117 & 337.46744 & +6.01526 & 4829 & 1.11 & -1.66 & -3 & -3.96$_{-0.44}^{+0.42}$ & 0.148 & 10.646 & 7.79$_{-1.37}^{+1.75}$ & -5.22 & 8.26 & 9.77 & -11.81\\
342809095 & 263.45129 & +18.10553 & 3870 & -0.24 & -2.17 & -22 & -6.12$_{-0.02}^{+0.02}$ & 0.053 & 11.280 & 29.47$_{-0.27}^{+0.27}$ & 12.49 & 21.36 & 24.74 & -13.47\\
318815093 & 233.31128 & +12.40625 & 4386 & 0.53 & -2.32 & 172 & -5.32$_{-0.16}^{+0.28}$ & 0.287 & 10.854 & 15.04$_{-1.82}^{+1.15}$ & 11.45 & 3.53 & 11.98 & -12.43\\
\hline\hline
\end{tabular*}
\end{table*}

In this section we select 5,171 halo-like RGB stars with [Fe/H]$<-1$ and $M_{Ks}<-4$ from the LAMOST DR3 catalog to map the stellar density profile for the stellar halo. The cut in $K_s$-band absolute magnitude ensures that the data is roughly complete within 50\,kpc from the Galactic center. The first three rows of the sample are listed in Table~\ref{tab:hRGB}. The columns in the table are explained in Table~\ref{tab:datamodel}. The full catalog can be found at \href{https://github.com/liuchaonaoc/LAMOST_density}{https://github.com/liuchaonaoc/LAMOST\_density}. We then derive the stellar densities at the spatial positions at which the halo-like RGB stars are located. 

Fig.~\ref{fig:nuHalo} shows the averaged stellar density map in the $R$--$Z$ plane for the halo population. The metal-poor RGB stars well probe the shape of the halo within a Galactiocentric radius of 50\,kpc. The contours of $\ln\nu=-12.5$ and $-13.5$ in the figure show clear oblate shapes within 20\,kpc. The iso-density contour at $-14.5$ displays a slightly larger axis ratio, although it is still oblate. Beyond 30\,kpc, the contours of $-15.5$ and $-16.5$ are roughly spherical. Although Xue et al. (\cite{xue2015}) claimed that a single power-law halo with variable axis ratio can better fit their data than the broken power-law,  it is probably for the first time, to our knowledge, that the variable axis ratio from inner to outer halo is directly illustrated in the  tomographic map. 

We further demonstrate how the broken power-law comes out with the assumption of a constant axis ratio. Fig.~\ref{fig:haloprofile} shows $\ln\nu$ as a function of $r$ for the points shown in Fig.~\ref{fig:nuHalo} with different fixed values of the axis ratio. The top panel presumes the axis ratio $q=1.0$. Thus $r$ is the same as the Galactocentric radius in spherical coordinates. The $\ln\nu$ profile is fitted with a broken power-law. Within $\sim40\pm2.8$\,kpc, the best fit power-law index is $-2.9\pm0.1$, while beyond this it is down to $-5.0\pm1.2$. In the bottom panel, $q$ is set to be $0.75$. In this case, the broken power-law profile shown in the top panel almost disappears. Instead, it shows that the two power-law indices in the inner and outer halo are similar, i.e., a single power-law may be better to fit the data. This means that the presumption of the axis ratio can significantly change the radial profile of $\ln\nu$. More quantitative study about the structure of the stellar halo with the LAMOST RGB samples can be found in Xu et al. (in preparation).

\section{Discusions}
\label{sect:discussion}
\subsection{The plate-wide smoothing density profile}

\begin{figure*}[htbp]
\hspace{-2cm}
	\begin{minipage}{18cm}
		\centering
		\includegraphics[scale=0.4]{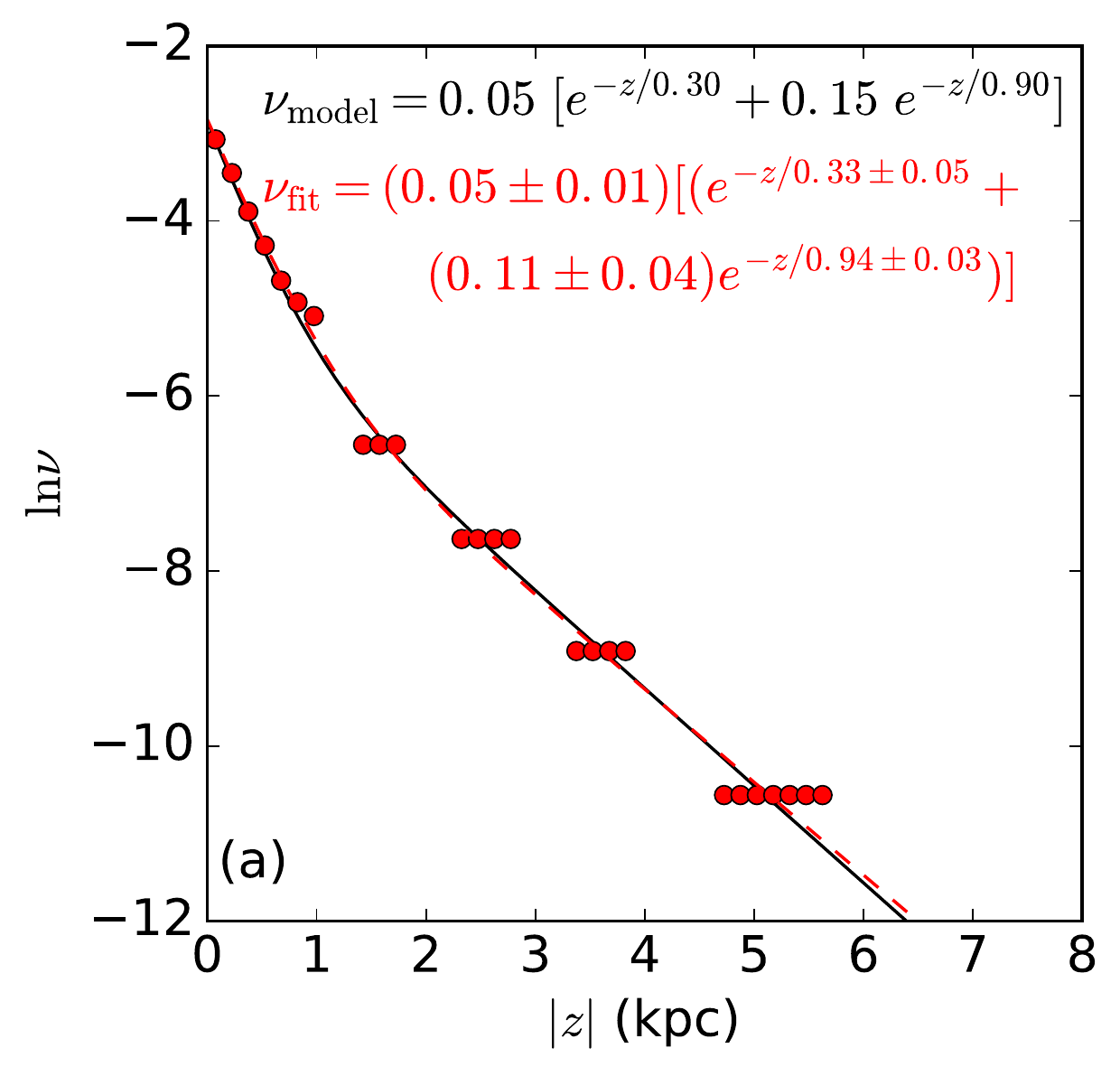}
		\includegraphics[scale=0.4]{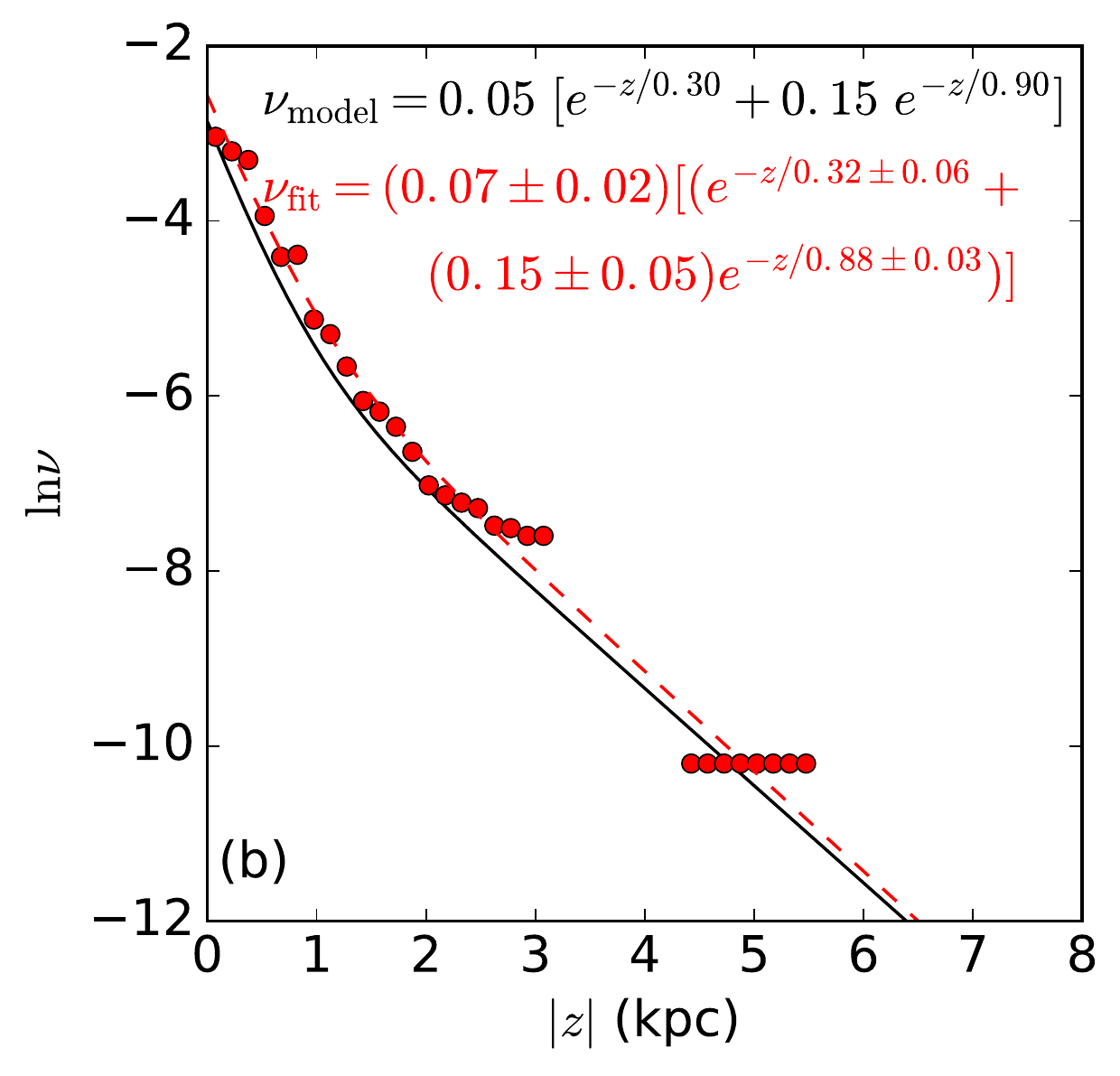}
		\includegraphics[scale=0.4]{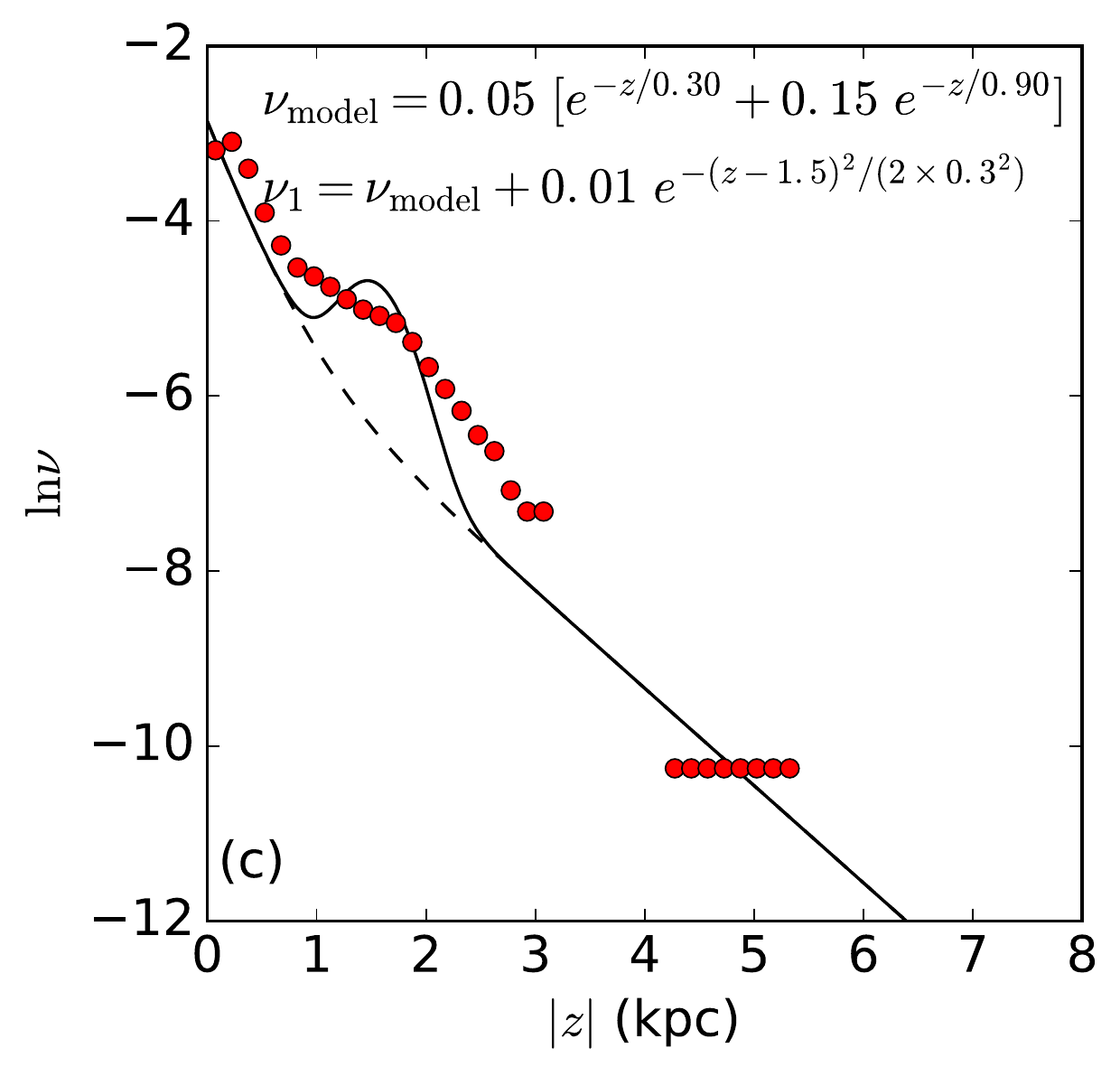}
	\end{minipage}
	\caption{Panel (a) shows the result of the vertical density profile estimates at $R=12$\,kpc for the toy model. The black solid line shows the profile from the toy model (see the details in text), while the red filled circles display the averaged stellar densities over the mock stars at various $Z$ bins. The red dashed line shows the best fit result for the red filled circles. The ``true'' parameters of the toy model are shown in the top line in the panel and the best fit parameters are shown in the second line. Panel (b) shows the similar result of the vertical density profile at $R=20$\,kpc. The symbols in this panel are same as Panel (a). Panel (c) shows similar result with an additional Gaussian-like substructure, which is centered at $z=1.5$\,kpc, superposed on the double-exponential toy model at $R=20$\,kpc. The first text line indicates the toy model of the double-exponential profile, which corresponds to the black dashed line, while the second text line shows the analytical form of the Gaussian-like substructure.}\label{fig:testplateeff}
\end{figure*}

In this work, we derive the density profile along a line-of-sight for each LAMOST plate. This implies that the stellar density estimate is smoothed over the solid angle of 20 square degrees, since the field-of-view of the LAMOST telescope is 5 degrees in diameter. Note that the stellar density profile may dramatically change from one end to another in such a large sky area, especially when the distance is large. As a consequence, the plate-wide smoothing density profile may smear out or blur the structure of the Milky Way to some extent. In this section we investigate the influence of the plate-wide smoothing to the density profile with toy models.

We adopt a toy model of the vertical stellar density profile with two exponential components: the thin disk component has the scale height of 0.3\,kpc and the thick disk has the scale height of 0.9\,kpc. At $Z=0$, the thick disk component is 15\% of the thin one. The normalizer, i.e. the stellar density at $Z=0$ is set to 0.05\,pc$^{-3}$. Such a configuration is similar to the Milky Way disk in the solar neighborhood.

Considering that the LAMOST plates usually overlap with each other in most of the sky areas, we draw mock stars from the toy model in the partly superposed LAMOST-like plates, i.e. circles with 5-degree-field-of-view, centered at $b=0^\circ$, $1^\circ$, $2^\circ$, $\dots$, $12^\circ$, $22^\circ$, $\dots$, $82^\circ$ with $l=180^\circ$. Obviously, these plates correspond to different vertical height above the Galactic mid-plane. The larger the distance, the larger the vertical height.

For each plate, at a given Galactocentric radius, we randomly draw 50 mock stars evenly distributed within the circle area of the plate. Note that, even though the mock stars have same radius, their $Z$ values are different because they are distributed at various Galactic latitudes within the plate. According to Section~\ref{sect:method}, the stellar densities at the positions of the mock stars are derived by averaging over the model density covering the area of the plate at the given radius. Thus, although the $Z$ values are different for the mock stars drawn from the same plate at the given radius, the stellar densities assigned to them are same. This is the similar but simplifed way to determine the stellar density as for the real observed data applied in Section~\ref{sect:density}. In spite that the uncertainties of the distance estimates and the selection function corrections are ignored, the simplied method to the stellar density determination is sufficient for investigating the effect induced by the plate-wide smoothing.

\subsubsection{The smoothing effect in the vertical disk profile}
We first investigate the effect in the vertical density profile. Fig.~\ref{fig:testplateeff} (a) shows the result of the vertical density profile test at $R=12$\,kpc. The black solid line indicates the ``true'' profile from the toy model. The red filled circles represent the averaged density values over the mock stars at different $Z$ bins. The red dashed line displays the best-fit double-exponential profile with the red symbols. It is seen that the derived stellar densities perfectly follow the ``true'' stellar density profile and the best-fit parameters are quite consistent with the ``true'' values with uncertainties less than 20\%.

We then move to larger distance at $R=20$\,kpc, where the effect of the plate-wide smoothing should be more substaential. The result in Fig.~\ref{fig:testplateeff} (b) shows that the shape of the best-fit density profile (red dashed line) derived from the ``observed'' data (red filled circles) is in agreement with the ``true'' profile (black solid line). Similar to the test at $R=12$\,kpc, the uncertainty of the derived model parameters is less than 20\% without substential systematic bias. In both cases at $R=12$ and $20$\,kpc, the plate-wide smoothing due to the large field-of-view does not significantly change the result.

\subsubsection{The smoothing effect in Monoceros ring-like substructure}
In section~\ref{sect:disk}, the well known Monoceros ring is not substentially exhibited in the outer Galactic disk, according to the stellar density profile estimates. Could it be possible that the smoothing in the large field-of-view of the LAMOST plate smears out such kind of substructure? We superpose an additional Gaussian-like substructure with the peak located at 1.5\,kpc and $\sigma$ of 0.3\,kpc to the toy model with a double-exponential vertical profile. The peak density of the Gaussian-like substructure is 20\% of the disk density at $Z=0$. The analytic form of the Gaussian-like substructure is indicated in the second text line in Fig.~\ref{fig:testplateeff} (c). The ``true'' vertical density profile with the Gaussian-like substructue located at $R=20$\,kpc is shown as the black solid line in Fig.~\ref{fig:testplateeff} (c). The red filled circles in the figure are the approximated density values obtained from the mock data. Compared with the derived best-fit stellar density in the panel (b) (red dashed line), it is seen that the approximated stellar density values do show a broader bump between $Z=1$ and $2$\,kpc, which corresponds to the Gaussian-like substructure. This means that even the substructure has a relatively smaller typical scale of 0.3\,kpc it can still be identified after the spatial smoothing induced by the large field-of-view, although the resulting substructure may be broadened to some extent. If the substructure is very weak, however, its signal may also be weakend by the broadening effect.

\subsubsection{Summary}
To summarize, the stellar density profile derived from the spatial averaging over the 5-degree-field-of-view may not significantly distort the derived Galactic structure. Meanwhile, although the smoothing with relatively large sky area may broaden the existing substructures, small-scale substructure with typcal scale of 0.3\,kpc can still be identified from the resulting stellartdensity profile. Therefore, the disappearing of the Monoceros ring in our result may not be due to the spatial smoothing induced by the applied technique.

\subsection{The \emph{post-observation} selection function}
The selection function used in this work is determined by comparison between the color-magnitude diagram from the photometric data and that from the final spectroscopic catalog. The selection function in the latter one is not only composed of the targeting selection, but also consider the selection effects induced during obserations and data reductions.%We do not need to know the originally designed selection function for the spectroscopic survey, but only need to ensure that the original selection was made based on the color index and magnitude. That is, no other quantity (e.g., metallicity, proper motion, age) is involved in the selection.}
The original selection function is always altered during the observations and data processing. For instance, one line-of-sight is designed to be observed multiple times with different ranges of magnitudes so that a larger range of magnitude can be covered. But due to the weather, some designed plates may never be observed. Then the \emph{post-observation} selection function for the line-of-sight would never be identical to the designed one. Another instance is that the data reduction may lose the spectra with very low signal-to-noise ratio, which may more frequently occur for fainter sources. Therefore, only using the originally designed targeting selectuion for the correction cannot take into account the effects from observations and data reductions.

\section{Conclusions}
\label{sect:conclusion}
In this paper, we introduce a statistical method to measure the stellar density profile along a given line-of-sight using spectroscopic survey data. This technique can flexiblly deal with the stellar density for different stellar populations. 

 %Consequently, we provide the way to derive the stellar density profile for any selected stellar population.

Our validation tests based on \emph{Galaxia} mock data demonstrate that we can well reconstruct the stellar density profile for a sub-population. Moreover, even if the sub-population contains quite few stars, we can still obtain reasonable density values at the spatial positions at which the stars are located. The tests show that in the worst cases, the derived stellar density may have a systematic bias of about $\Delta\nu/\nu\sim1$. However, large surveys like LAMOST can observe many plates covering a wide area of the sky. Thus, statistically, the final averaged stellar density in the small discrepant regions can reduce or cancel the different systematic biases occurring in different lines-of-sight, and finally achieve relatively high precision when averaged over large areas.

Finally, we apply the method to the LAMOST DR3 RGB stars. For the Galactic outer disk, we find that 1) the disk component still contributes to the total stellar surface density by about 10\% at $R=19$\,kpc, implying that our Galaxy has a quite large stellar disk; 2) we do not observe any significant truncation of the disk, but only see a smooth transition from the disk to the halo at around $20$\,kpc; 3) the Monoceros ring is not seen in our density map of the outer disk; and 4) the disk seems vertically asymmetric beyond $R\sim12$\,kpc.

We confirm that the stellar halo is oblate in the inner halo (within $r\sim30$\,kpc) and becomes spherical in the outer part. If we presume a constant axis ratio of 1.0, the stellar density profile shows a clear broken power-low similar to some previous works (e.g., Deason et al. \cite{deason2011}). However, if we change the constant axis ratio to 0.75, the broken power-law disappears. This means that whether the halo profile is broken or not depends on the axis ratio. The tomographic map of the stellar halo displayed in this work shows evidence that the halo has a radially varying axis ratio. Therefore, either using a single or broken double power-law with constant axis ratio does not properly reflect the real structure of the stellar halo.

%More quantitative analysis about the disk and halo will be discussed in Wang et al. in preparation and Xu et al. in preparation, respectively.

\begin{acknowledgements}
We thank the anonymous referee for the very helpful comments. We thank Li Chen and Jing Zhong for helpful discussions. This work is supported by the Strategic Priority Research Program ``The Emergence of Cosmological Structures" of the Chinese Academy of Sciences, Grant No. XDB09000000 and the National Key Basic Research Program of China 2014CB845700. C. L. acknowledges the NSFC under grants 11373032 and 11333003. Guoshoujing Telescope (the Large Sky Area Multi-Object Fiber Spectroscopic Telescope LAMOST) is a National Major Scientific Project built by the Chinese Academy of Sciences. Funding for the project has been provided by the National Development and Reform Commission. LAMOST is operated and managed by the National Astronomical Observatories, Chinese Academy of Sciences.
\end{acknowledgements}
%\appendix                  %%appendicial material is supported

\clearpage

\appendix
\section{Test the method with simulation data at $l=180^\circ$ and $b=0^\circ$}
\begin{figure*}[htbp]
\centering
\begin{minipage}{18cm}
\centering
\includegraphics[scale=0.4]{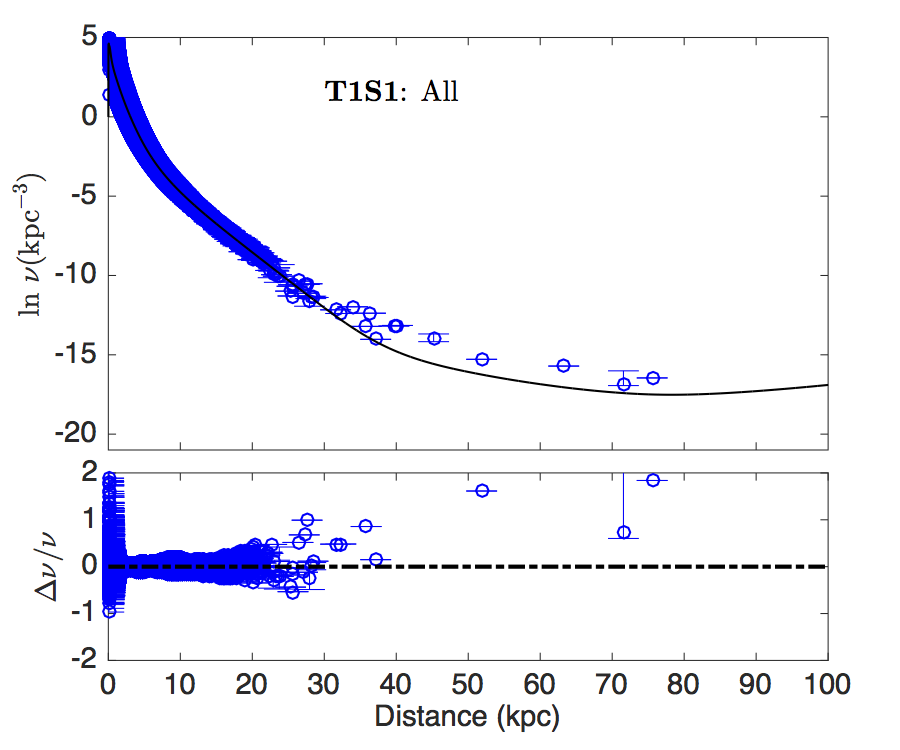}
\includegraphics[scale=0.4]{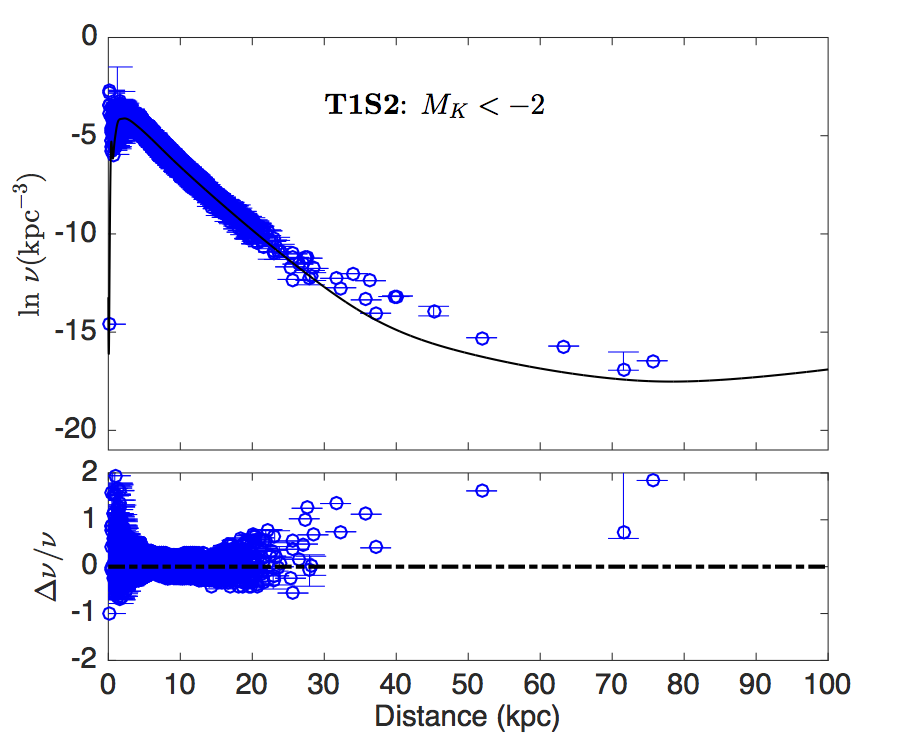}
\includegraphics[scale=0.4]{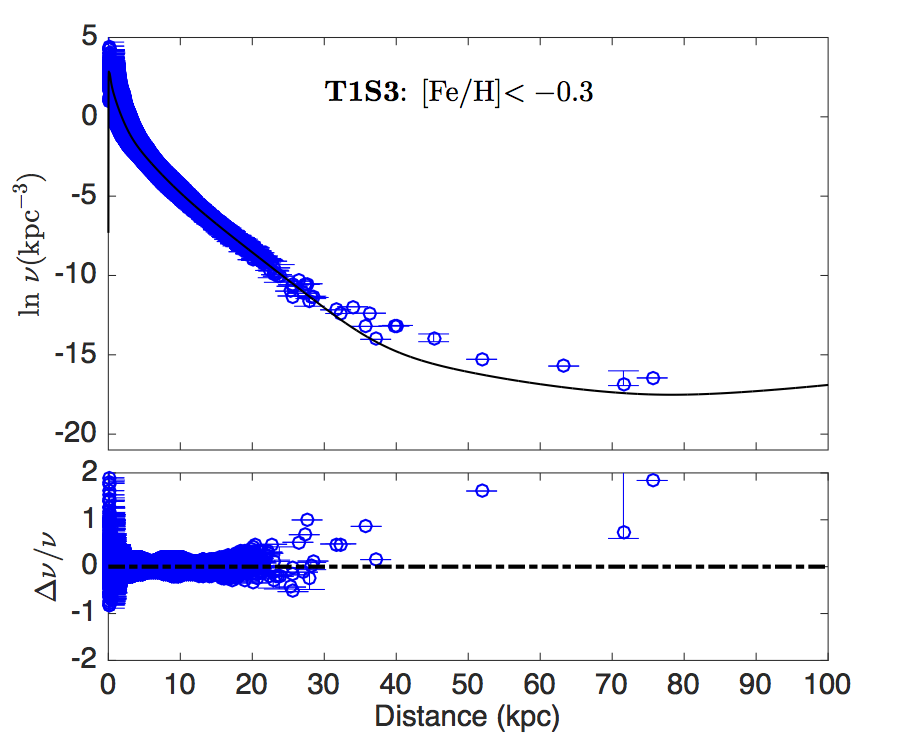}
\includegraphics[scale=0.4]{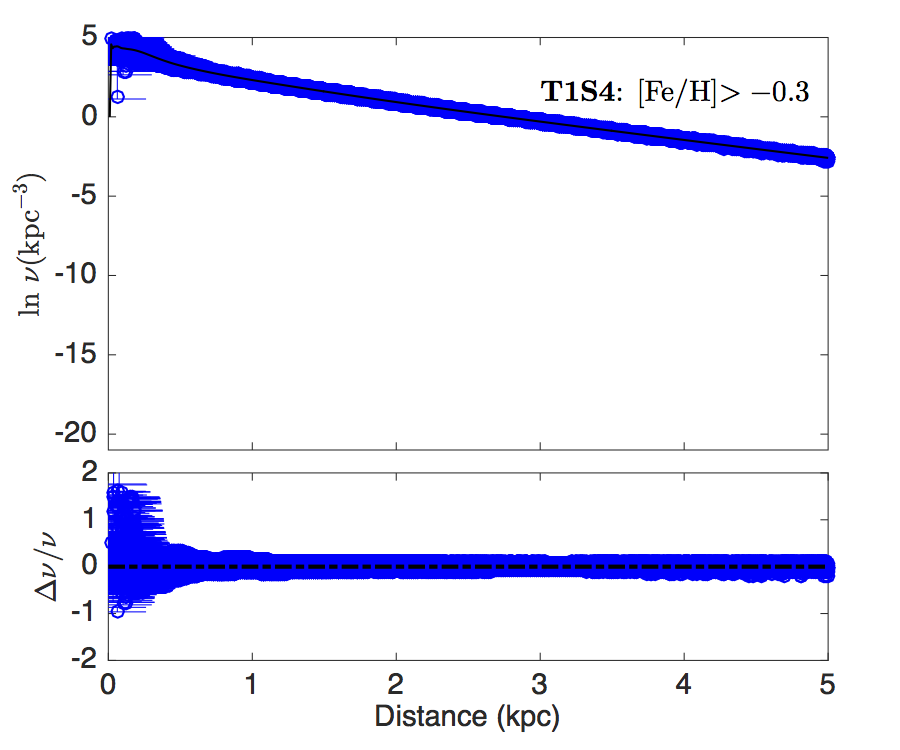}
\end{minipage}
\caption{Similar to Fig.~\ref{fig:nu1}, but with the selection function \textbf{T1} for the \emph{Galaxia} simulation data located at $l=180^\circ$ and $b=0^\circ$.}\label{fig:nu3}
\end{figure*}

\begin{figure*}[htbp]
\centering
\begin{minipage}{18cm}
\centering
\includegraphics[scale=0.4]{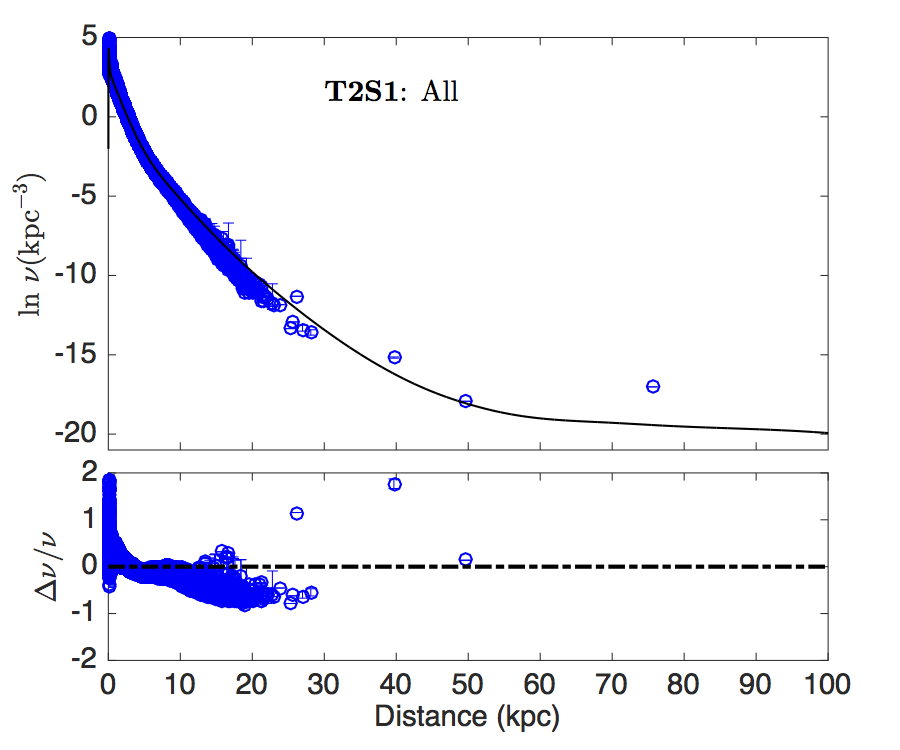}
\includegraphics[scale=0.4]{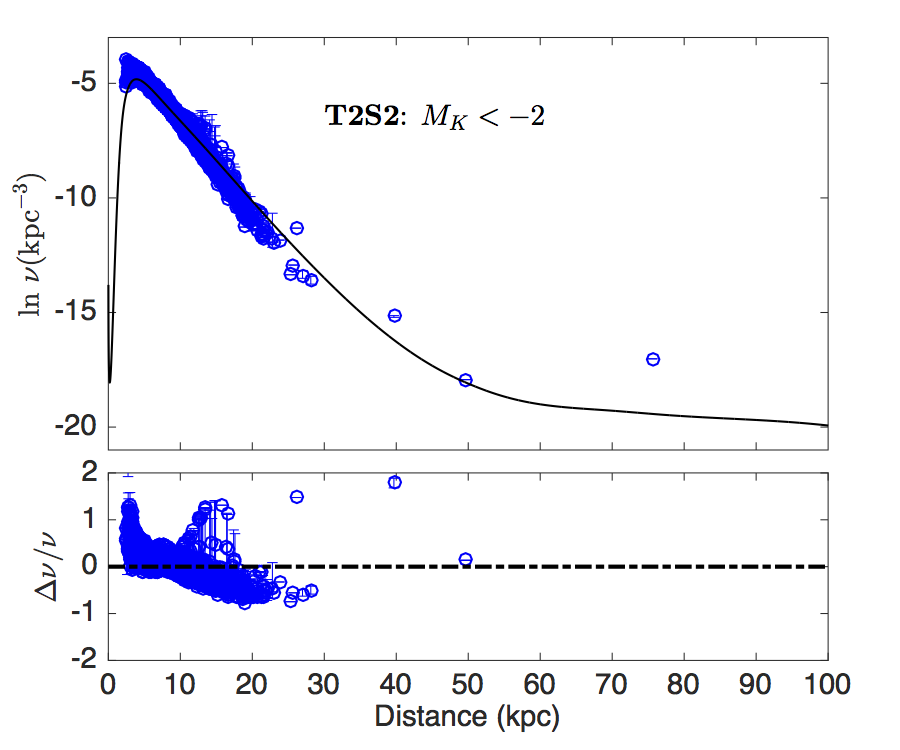}
\includegraphics[scale=0.4]{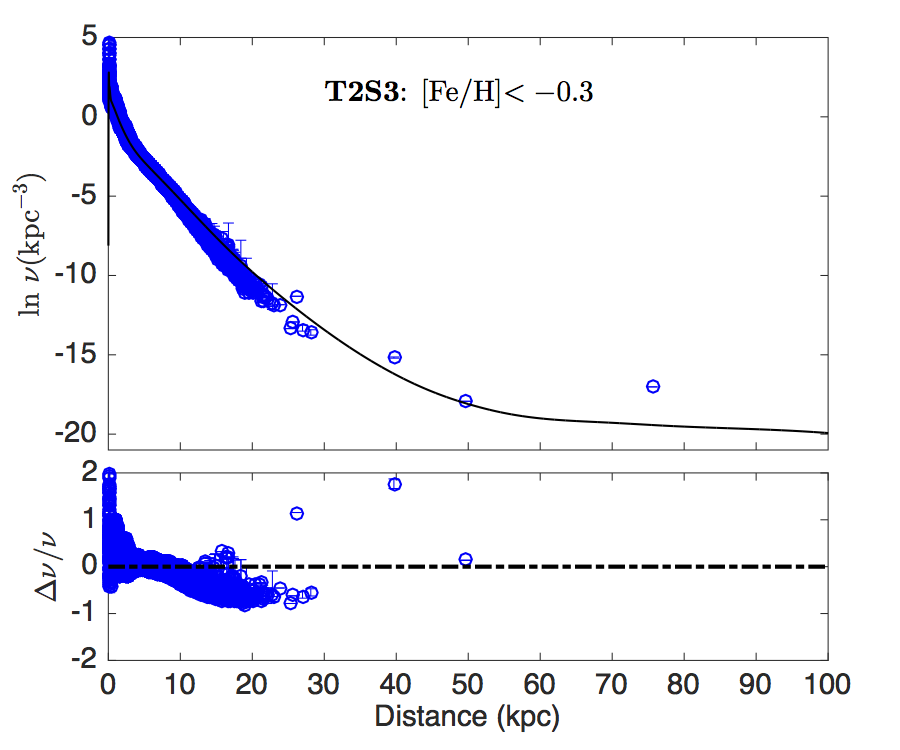}
\includegraphics[scale=0.4]{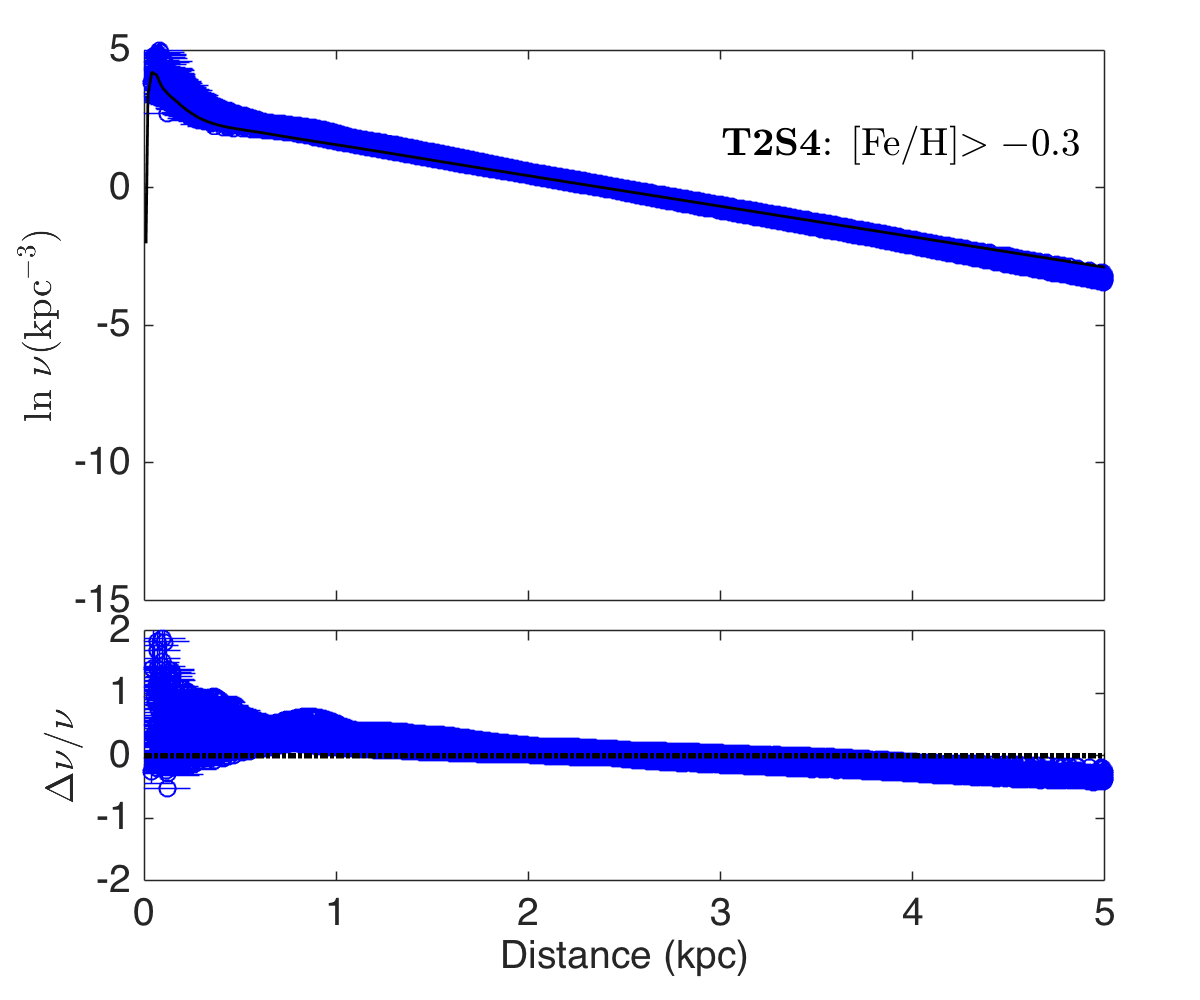}
\end{minipage}
\caption{Similar to Fig.~\ref{fig:nu1}, but with the selection function \textbf{T2} for the \emph{Galaxia} simulation data located at $l=180^\circ$ and $b=0^\circ$.}\label{fig:nu4}
\end{figure*}

In this section we further provide an additional test for the method using the \emph{Galaxia} simulation data located at $l=180^\circ$ and $b=0^\circ$, at which the extinction is significantly large, with field of view of 20 squared degrees. Fig.~\ref{fig:nu3} shows the test result with the selection function {\textbf T1}, while Fig.~\ref{fig:nu4} shows the test result with the selection function {\textbf T2}. Under both selection functions, the performance of the stellar density estimates for the stars with higher extinction do not show significant difference with the previous tests, as shown in Figs.~\ref{fig:nu1} and~\ref{fig:nu2}.

\label{lastpage}

\end{document}